\documentclass[aps,prl,preprintnumbers,twocolumn,superscriptaddress,nofootinbib,floatfix,10pt]{revtex4-1}
\usepackage{amsfonts,amssymb,stmaryrd,latexsym,amsmath,braket}
\usepackage{graphicx,subfigure}
\usepackage{times}
\usepackage{slashed}
\usepackage{braket}
\usepackage{verbatim}
\usepackage[utf8]{inputenc}
\newcommand{\beginsupplement}{%
        \setcounter{table}{0}
        \renewcommand{\thetable}{S\arabic{table}}%
        \setcounter{figure}{0}
        \renewcommand{\thefigure}{S\arabic{figure}}%
        \setcounter{equation}{0}
        \renewcommand{\theequation}{S\arabic{equation}}        
     }

\usepackage{color}
\newcommand{\hh}[1]{{\color{red}[HH: #1]}}

\begin{document}

\title{Hidden spin-isospin exchange symmetry}

\author{Dean Lee}

\affiliation{Facility for Rare Isotope Beams and Department of Physics and Astronomy,
Michigan State University, MI 48824, USA}

\author{Scott Bogner}

\affiliation{Facility for Rare Isotope Beams and Department of Physics and Astronomy,
Michigan State University, MI 48824, USA}

\author{B. Alex Brown}

\affiliation{Facility for Rare Isotope Beams and Department of Physics and Astronomy,
Michigan State University, MI 48824, USA}

\author{Serdar Elhatisari}

\affiliation{Faculty of Engineering, Karamanoglu Mehmetbey University, Karaman
70100, Turkey}

\author{Evgeny Epelbaum}

\affiliation{Ruhr-Universit\"at Bochum, Fakult\"at f\"ur Physik und Astronomie,
Institut f\"ur Theoretische Physik II,
D-44780 Bochum, Germany}

\author{Heiko Hergert}

\affiliation{Facility for Rare Isotope Beams and Department of Physics and Astronomy,
Michigan State University, MI 48824, USA}

\author{Morten Hjorth-Jensen}

\affiliation{Facility for Rare Isotope Beams and Department of Physics and Astronomy,
Michigan State University, MI 48824, USA}
\affiliation{Department of Physics and Center for Computing in Science Education, University of Oslo, N-0316 Oslo, Norway}

\author{Hermann Krebs}

\affiliation{Ruhr-Universit\"at Bochum, Fakult\"at f\"ur Physik und Astronomie,
Institut f\"ur Theoretische Physik II,
D-44780 Bochum, Germany}

\author{Ning Li}

\affiliation{Sun Yat-sen University, School of Physics, 135 Xingang Road, Haizhu District,
Guangzhou, Guangdong Province, 510000, China}

\author{Bing-Nan Lu}

\affiliation{China Academy of Engineering Physics, Graduate School, Building 8, No.~10 Xi'er Road, ZPark II, Haidian District, Beijing, 100193, China}

\author{Ulf-G. Mei{\ss}ner}

\affiliation{Helmholtz-Institut f\"ur Strahlen- und Kernphysik and Bethe Center for Theoretical Physics, Universit\"at Bonn, D-53115 Bonn, Germany}

\affiliation{Institute for Advanced Simulation, Institut f\"ur Kernphysik, and~J\"ulich~Center for Hadron Physics, Forschungszentrum J\"ulich,
D-52425 J\"ulich, Germany}

\affiliation{Tbilisi State University, 0186 Tbilisi, Georgia}

\begin{abstract}
The strong interactions among nucleons have an approximate spin-isospin exchange symmetry that arises from the properties of quantum chromodynamics in the limit of many colors, $N_c$.   However this large-$N_c$ symmetry is well hidden and reveals itself only when averaging over intrinsic spin orientations. Furthermore, the symmetry is obscured unless the momentum resolution scale is close to an optimal scale that we call $\Lambda_{{\rm large-}N_c}$.  We show that the large-$N_c$ derivation requires a momentum resolution scale of $\Lambda_{{\rm large-}N_c} \sim 500$~MeV.  We derive a set of spin-isospin exchange sum rules and discuss implications for the spectrum of $^{30}$P and applications to nuclear forces, nuclear structure calculations, and three-nucleon interactions.
\end{abstract}


\maketitle

Quantum chromodynamics is the fundamental theory of the strong interactions.  While the physical universe has $N_c=3$ colors, it is useful to consider how nuclear physics might appear when $N_c$ is large \cite{tHooft:1973alw,Witten:1979kh,Muther:1987sr,Dashen:1993as,Dashen:1994qi,Carone:1993dz,Jenkins:1994md}.  In the large-$N_c$ picture, Kaplan, Savage, and Manohar \cite{Kaplan:1995yg,Kaplan:1996rk} found that the leading nucleon-nucleon (NN) interaction between two nucleons can be written in the form
\begin{equation}
V_{{\rm large-}N_c}^{\rm 2N} =     V_C + \vec{\sigma}_1\cdot \vec{\sigma}_2 \vec{\tau}_1 \cdot \vec{\tau}_2 W_S + S_{12} \vec{\tau}_1 \cdot \vec{\tau}_2 W_T + \ldots, \label{largeNc}
\end{equation}
where the ellipses refer to terms subleading in the large-$N_c$ expansion, $\vec{\sigma}_{1,2}$ represent the Pauli matrices for intrinsic spins, $\vec{\tau}_{1,2}$ are the Pauli matrices for isospin, and $S_{12}$ is the tensor operator $3\hat{r} \cdot \vec{\sigma}_1 \hat{r} \cdot \vec{\sigma}_2 -\vec{\sigma}_1 \cdot \vec{\sigma}_2$.  
Each of the scalar functions $V_C$, $W_S$ and $W_T$ are local interactions, meaning that they depend only on the separation vector between the two nucleons and not their velocities.  The strength of these leading interactions scale linearly with $N_c$, while all other terms scale as $1/N_c$ or smaller.  The relative $1/N_c^2$ suppression of the subleading terms is fairly strong even for $N_c=3$.  

In the literature, the large-$N_c$ limit is often linked with Wigner's approximate SU(4) spin-isospin symmetry \cite{Wigner:1937} where the four spin and isospin nucleon components transform as an SU(4) quartet.  See for example, Ref.~\cite{Kaplan:1995yg,Mehen:1999qs,Chen:2004rq,Lee:2007eu,CalleCordon:2009ps,Konig:2016utl,Lu:2018bat}.  In the low-energy limit, nucleon-nucleon scattering is dominated by the two S-wave channels.  Therefore, if one reduces Eq.~(\ref{largeNc}) to only the S-wave channels, then the tensor interaction vanishes and we can replace $\vec{\sigma}_1\cdot \vec{\sigma}_2 \vec{\tau}_1 \cdot \vec{\tau}_2$ by $-3$.  We therefore seem to derive the result that the low-energy S-wave interactions satisfy Wigner's SU(4) symmetry.  On the other hand, the deuteron is bound by $2.22$ MeV while the spin-singlet channel is unbound. 
The discrepancy seems larger than the predicted $1/N_c^2$ relative error of the large-$N_c$ expansion.

As we will show in the following derivation, Eq.~(\ref{largeNc}) is only valid when the momentum resolution scale is not too high or too low (where the precise meaning of high and low will become clear later).  At the proper resolution scale we find that the large-$N_c$ limit predicts a spin-isospin exchange symmetry that is satisfied at the $1/N_c^2$ level and provides a valuable guide for understanding nuclear interactions and nuclear structure.

The derivation of Eq.~\eqref{largeNc} can be sketched as follows. Let us assume that $N_c$ is a large odd integer, $N_c=2k+1$.  The nucleon wave function in its rest frame has a fairly simple structure in the limit of non-relativistic quarks \cite{Witten:1979kh,Karl:1984cz}.  The $N_c$ quarks are completely antisymmetric with respect to color, and the spatial wave function is a simple Hartree product state.  The spin and isospin content can be understood by dividing the quarks into two partitions of size $k+1$ and $k$.  The partition with $k+1$ quarks has isospin $(k+1)/2$ and spin $(k+1)/2$.  The partition with $k$ quarks has isospin $k/2$ and spin $k/2$.  These maximal values for isospin and spin mean that the wave functions for each partition are symmetric in both isospin and spin indices.  The two partitions are combined together to have total isospin $1/2$ and total spin $1/2$, and we symmetrize over all ways to divide the quarks into the two partitions.  The nucleon wave function has a spin-isospin exchange symmetry, which we can define at the quark level as the exchange of spin-up down quarks $d_{\uparrow}$ with spin-down up quarks $u_{\downarrow}$.  At the nucleon level, the spin-isospin exchange symmetry corresponds with the exchange of spin-up neutrons $n_{\uparrow}$ with spin-down protons $p_{\downarrow}$. 

If we now relax the condition that the quarks are non-relativistic, the generalization to the relativistic mean-field picture becomes more complicated.  However the phenomenological success of the non-relativistic quark model in predicting the large isovector part of the nucleon anomalous magnetic moment and nearly vanishing isoscalar part \cite{Mergell:1995bf,Belushkin:2006qa} suggest that the quantitative features of this description of the nucleon wave function remain valid.  We note that there have been lattice QCD studies of baryon magnetic moments \cite{Shanahan:2014uka,Parreno:2016fwu} and spin-flavor symmetries \cite{Wagman:2017tmp,Illa:2020nsi}.

The expectation value of the quark bilinear operator $\bar{q}q$ for the nucleon will scale linearly with the number quarks in the nucleon and thus linearly with $N_c$.  Similarly an operator such as $\bar{q}\sigma^a \tau^b q$ will also scale linearly with $N_c$.  However, since the spin of the nucleon is $1/2$ and the isospin is $1/2$, the expectation values of $\bar{q}\vec{\sigma}q$ and $\bar{q}\vec{\tau}q$ do not grow with $N_c$.  This argument no longer holds for $\bar{q}\vec{\sigma}q$ if the nucleon is moving with speed, $v>0$.  In that case the Pauli-Lubanski pseudovector $W_\mu$ will have a nonzero time component that scales as $v$ times $N_c$.

Let us now consider two nucleons with relative velocities low enough so that we have an approximate Galilean invariance for the nucleon wave functions.  Following Ref.~\cite{Kaplan:1996rk}, we can apply a multipole expansion to the effective nucleon-nucleon interactions that we can build from the leading quark operators $\bar{q}q$ and $\bar{q}\sigma^a \tau^b q$. That process leads to an effective Hamiltonian where the leading terms in the large-$N_c$ limit are given in Eq.~\eqref{largeNc}.  This construction yields local interactions, and the connection with meson exchange interactions has been studied \cite{Banerjee:2001js,Cohen:2002im} as well as predictions for scattering observables \cite{Cohen:2002qn}.

We note that Eq.~\eqref{largeNc} is not renormalization group invariant, and we are somehow implicitly setting a preferred momentum resolution scale. How this happens can be understood as follows. Our simple description of the nucleon wave function has corrections proportional to the square of the nucleon velocity, $v^2$.  We must control these high-energy modes in our effective theory in order for the derivation of Eq.~\eqref{largeNc} to hold with errors of size $1/N_c^2$.  In order to restrict $v^2$ so that it is of size $1/N_c^2$ or smaller, we require that the momentum resolution scale $\Lambda$ is proportional to the nucleon mass times a factor of $1/N_c$. Since the nucleon mass is proportional to $N_c$, the upper bound on $\Lambda$ remains a constant independent of $N_c$.  This is consistent with the discussion of nucleon momenta in Ref.~\cite{Banerjee:2001js}.  We discuss the relevant energy and momentum scales further in the Supplemental Material.

While there are no problems with nucleon-nucleon scattering at low energies, the form of Eq.~\eqref{largeNc} will no longer hold if we also choose to lower the momentum resolution scale $\Lambda$ so much that the distinction between the orbital angular momentum and intrinsic spin of the nucleon is not fully resolved.  In that case the nucleon intrinsic spin should be viewed as an effective spin composed of both intrinsic spin and orbital angular momentum.  In order to suppress these effects, it suffices that $1/\Lambda$ is comparable to the size of the nucleon or smaller.  Since the size of the nucleon is independent of $N_c$, the resulting lower bound on $\Lambda$ is also a constant independent of $N_c$.

We note that a similar phenomenon of hidden symmetry at low energies can also be seen in nucleon-nucleon scattering within the KSW scheme introduced in Ref.~\cite{Kaplan:1998tg,Kaplan:1998we}.  In that case the difference between the two S-wave spin channels is magnified at low energies due to the different scattering lengths, one large and positive and the other large and negative.

The renormalization group dependence of the large-$N_c$ approximation was first studied in Ref.~\cite{Timoteo:2011tt,Arriola:2013nja,RuizArriola:2016vap}.  The authors noticed that there was a momentum resolution scale at which the large-$N_c$ constraints are satisfied far better than at other scales and speculated that it might be a numerical accident.  The renormalization group analysis for pionless effective field was also considered in Ref.~\cite{Schindler:2018irz}.  See also Ref.~\cite{Beane:2018oxh} for a discussion of emergent symmetries as arising from entanglement suppression.  In this letter we make the stronger statement that the leading large-$N_c$ reduction of the nuclear interaction has error corrections of size $1/N_c^2$ only when the momentum resolution scale is near an optimal resolution scale that we call $\Lambda_{{\rm large-}N_c}$.
As we will see in the examples below, the optimal momentum resolution scale is 
$\Lambda_{{\rm large-}N_c}\sim 500$~MeV, consistent with the momentum scale found in Ref.~\cite{Timoteo:2011tt}. We posit that any effective field theory description of the nuclear interactions at momentum resolution scale $\Lambda_{{\rm large-}N_c}$ with local interactions will satisfy Eq.~\eqref{largeNc} with corrections of size $1/N_c^2$.  For interactions with a small amount of velocity dependence and therefore nonlocality, there will also be small corrections arising from the velocity dependence of the interactions.  For some explicit examples demonstrating this behavior, see Fig.~1 of Ref.~\cite{Gezerlis:2014zia} for local chiral potentials at N$^2$LO and Figs.~6,7 of Ref.~\cite{Reinert:2017usi} for the long-range part of chiral potentials up to N$^4$LO.

As an example, in Table~\ref{S} we show the S-wave short-range interaction coefficients in lattice units (l.u.) for lattice chiral effective field theory with lattice spacing $a=1.32$~fm.  This corresponds to a momentum cutoff of $\pi/a = 471$~MeV.  Each of these terms are central interactions, meaning that the spin components form rotational invariants through contractions with themselves.  We note the very good agreement between the $S=0,T=1$ and $S=1,T=0$ channels, where $S$ is intrinsic spin and $T$ is isospin.  The error bars indicate uncertainties in the fit to empirical data.   The full next-to-next-to-next-to-leading-order (N$^3$LO) chiral interaction includes these short-range interactions together with a leading-order (LO) interaction composed of an SU(4) symmetric interaction and one-pion exchange potential that together have the form described in Eq.~\eqref{largeNc}.  Full details are given in the Supplemental Materials.  In the last row of Table~\ref{S} we show the D-wave short-range interaction coefficients in lattice units.  In this case the interactions are not purely central, and so we average over all possible total angular momentum channels $J$ for the spin-triplet channels to remove contributions from the tensor force and spin-orbit interactions, and this is denoted as $^3$D$_{\rm all}$.   This averaging over $J$ was also used in Ref.~\cite{CalleCordon:2009ps}.  We see again very good agreement between the $S=0,T=1$ channel and the average over the $S=1,T=0$ channels.

\begin{table}
\caption{
Lattice chiral effective field theory coefficients of short-range interactions for the spin-singlet and spin-triplet S-wave and D-wave channels at lattice spacing $a=1.32$~fm. 
\label{S}
}
\centering{}%
\begin{tabular}{c c|c c}
\hline
channel, order & coupling (l.u.) & 
channel, order & coupling (l.u.) \\
\hline
$^1{\rm S}_0,Q^0$ & 1.45(5)  &
$^3{\rm S}_1,Q^0$ & 1.56(3) \\
$^1{\rm S}_0,Q^2$ & $-$0.47(3) & 
$^3{\rm S}_1,Q^2$ & $-$0.53(1) \\
$^1{\rm S}_0,Q^4$ & 0.129(13) &
$^3{\rm S}_1,Q^4$ & 0.115(4) \\
$^1{\rm D}_2,Q^4$ & $-$0.088(1)  &
$^3{\rm D}_{\rm all},Q^4$ & $-$0.070(2) \\
\hline
\end{tabular}
\end{table}

Let us now consider the matrix element between any two-nucleon states $A$ and $B$, both with total intrinsic spin $S$ and total isospin $T$.  Let $H$ be the isospin-invariant part of the nucleon-nucleon Hamiltonian. We define the matrix element $M(S,T)$ as
\begin{align} 
\frac{1}{2S+1}\sum_{S_z=-S}^S \braket{A;S,S_z;T,T_z|H|B;S,S_z;T,T_z}.
\label{M_AB}
\end{align}
The statement of spin-isospin exchange symmetry is the constraint $M(S,T)=M(T,S)$.  

As an example, let us consider the structure of $^{30}$P. In a minimal shell model description, $^{30}$P has one proton and one neutron in the $1s_{1/2}$ orbitals.  The actual wave function is considerably more complicated than this, but we can make the rough approximation that the spatial wave functions for the two lowest-lying states of $^{30}$P are the same. If true, then the prediction of spin-isospin exchange symmetry is that the $S=0,T=1$ and $S=1,T=0$ states are degenerate.  The actual data is that the $1^+$ ground state is lower than the $0^+$ state by about 0.677 MeV.  This is a fairly good agreement, far better than the splitting between the deuteron and the spin-singlet proton-neutron pair in vacuum.  In that case the tensor force significantly modifies the deuteron wave function.  Our analysis here suggests that the interactions of the proton-neutron pair with the $^{28}$Si core is suppressing spatial correlations of the $1^+$ wave function caused by the tensor interaction.

We can also look at the two-body matrix elements for the proton-neutron pair in the $1s_{1/2}$ orbitals. For the following calculations we use a harmonic oscillator frequency of $13.92$~MeV.   Starting from the AV18 potential \cite{Wiringa:1994wb} and using a renormalization group flow to construct the corresponding effective interaction $V_{{\rm low} k}$ \cite{Bogner:1999my} at momentum scale $\Lambda = 2.5~{\rm fm}^{-1}=490~{\rm MeV}$, we get  two-body matrix matrix elements of $-2.54$~MeV for the $1^+$ state and $-2.37$~MeV for the $0^+$ state, 
respectively, showing a relative error consistent with the expected $1/N_c^2$ size.  Using the same $V_{{\rm low} k}$ renormalization group flow for the chiral N$^3$LO interaction \cite{Entem:2003ft} at $\Lambda = 2.5~{\rm fm}^{-1}$, we get similar results of $-2.48$~MeV for the $1^+$ state and $-2.36$~MeV for the $0^+$ state.
However, the spin-isospin exchange symmetry is obscured at other resolution scales.  For AV18 at $\Lambda = 6.0~{\rm fm}^{-1}$ we find $1.58$~MeV for the $1^+$ state and $0.19$~MeV for the $0^+$ state, which is a violation larger than the expected $1/N_c^2$ size.  This lends support to our argument that the large-$N_c$ derivation is only valid for momentum resolution scales in the range $\Lambda_{{\rm large-}N_c}\sim 500$~MeV.  We note that the chiral N$^3$LO interaction has some nonlocality while the AV18 is a local interaction.  In the Supplemental Materials we consider several other interactions and check for spin-isospin exchange symmetry as measured by scattering phase shifts.

Our test of spin-isospin exchange symmetry is sensitive to the effective resolution scale of the nuclear interactions.  We can therefore use spin-isospin exchange symmetry as a tool for estimating the effective resolution scale $\Lambda$ of any nuclear interaction which might use one of many possible methods for regulating high-energy behavior.  For $\Lambda < \Lambda_{{\rm large-}N_c}$ the $(S,T) = (1,0)$ central interaction is more attractive, and for $\Lambda > \Lambda_{{\rm large-}N_c}$ the $(S,T) = (0,1)$ central interaction is more attractive.  We can understand this as follows.  As we lower $\Lambda$, the effect of the tensor coupling between the S and D partial waves becomes weaker.  In order to compensate for the loss of tensor attraction, the $(S,T) = (1,0)$ central interaction must be more attractive than the $(S,T) = (0,1)$ central interaction.  As we increase $\Lambda$, the tensor attraction becomes stronger and the reverse effect occurs.

\begin{figure*}[ht]
\centering
\includegraphics[width=4cm]{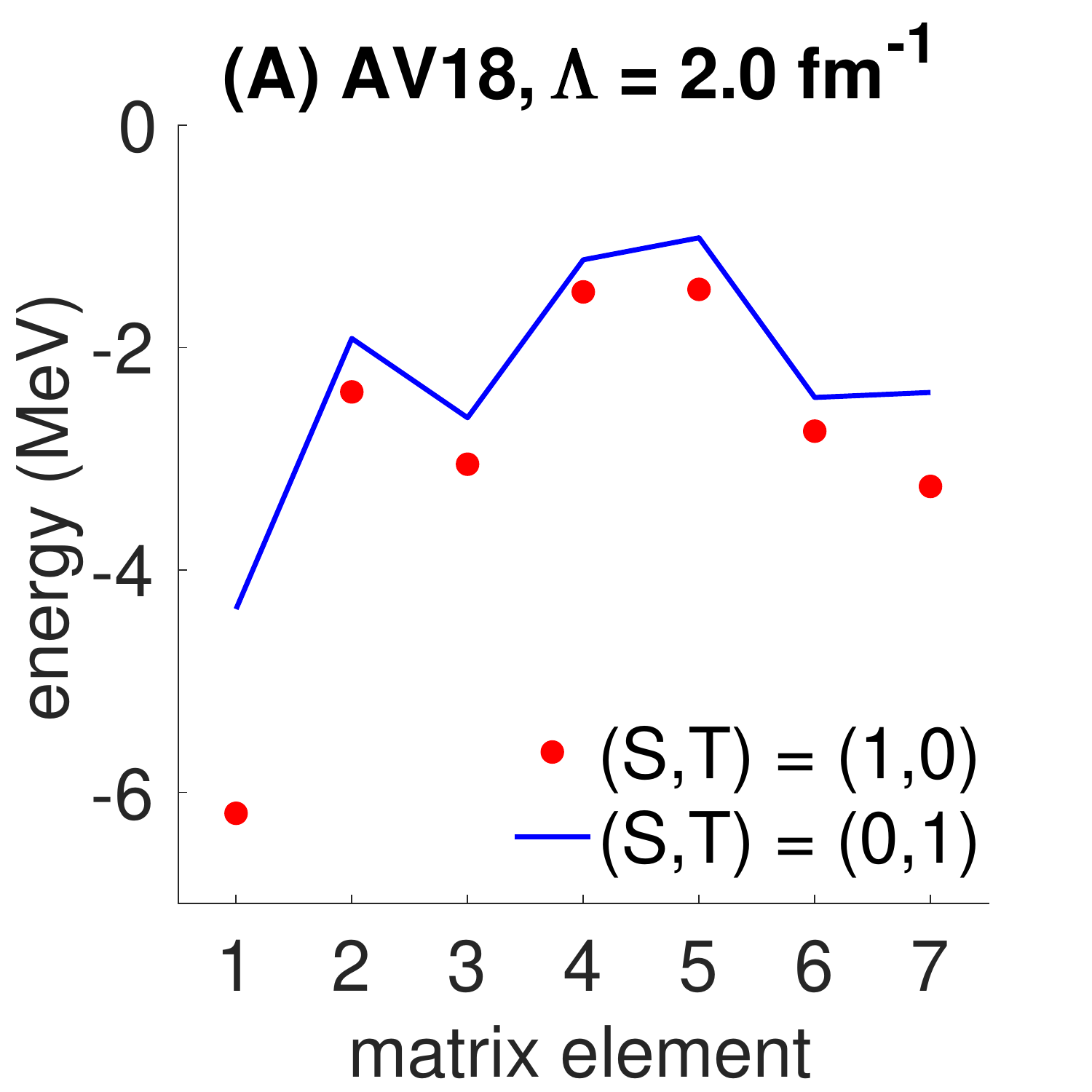}
\includegraphics[width=4cm]{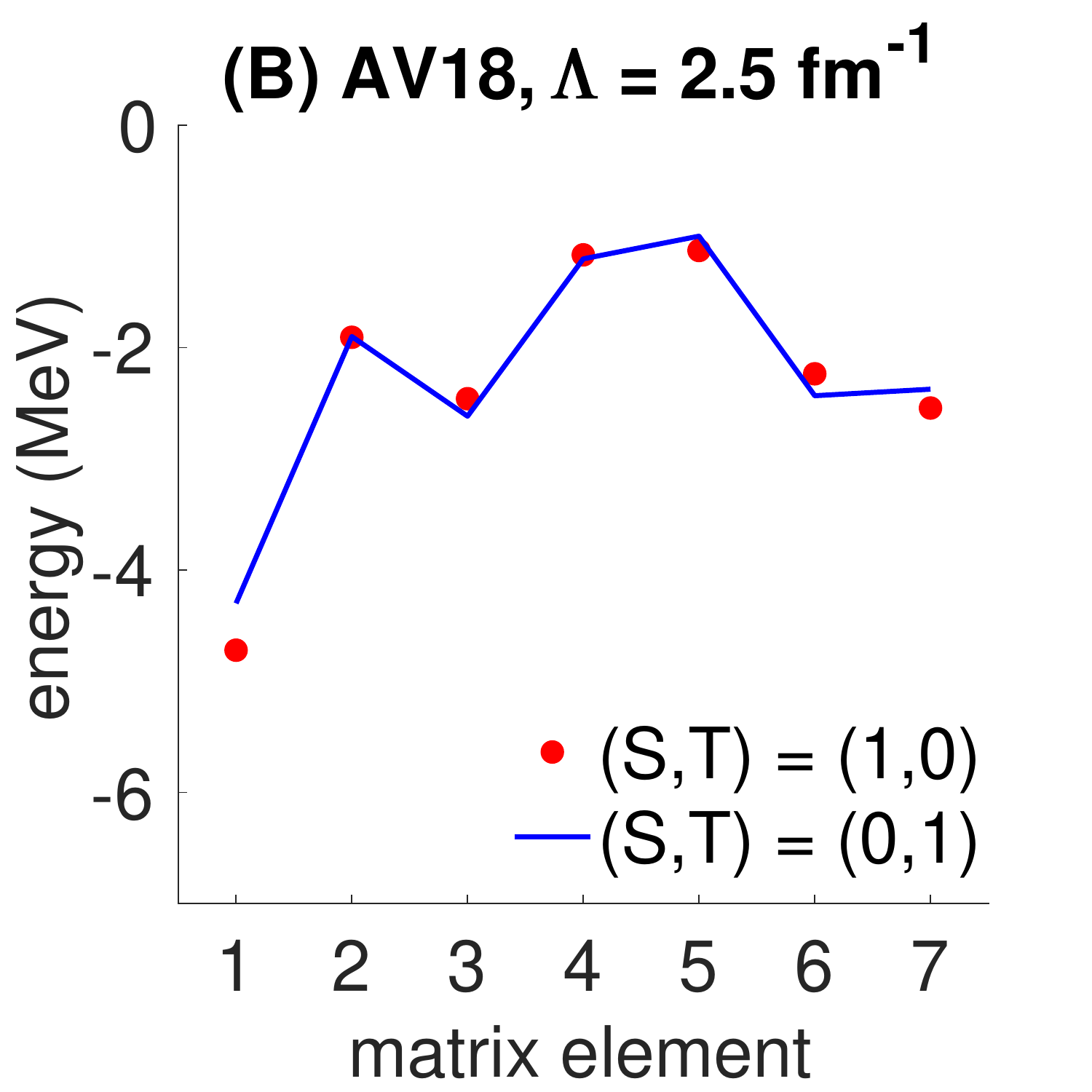}
\includegraphics[width=4cm]{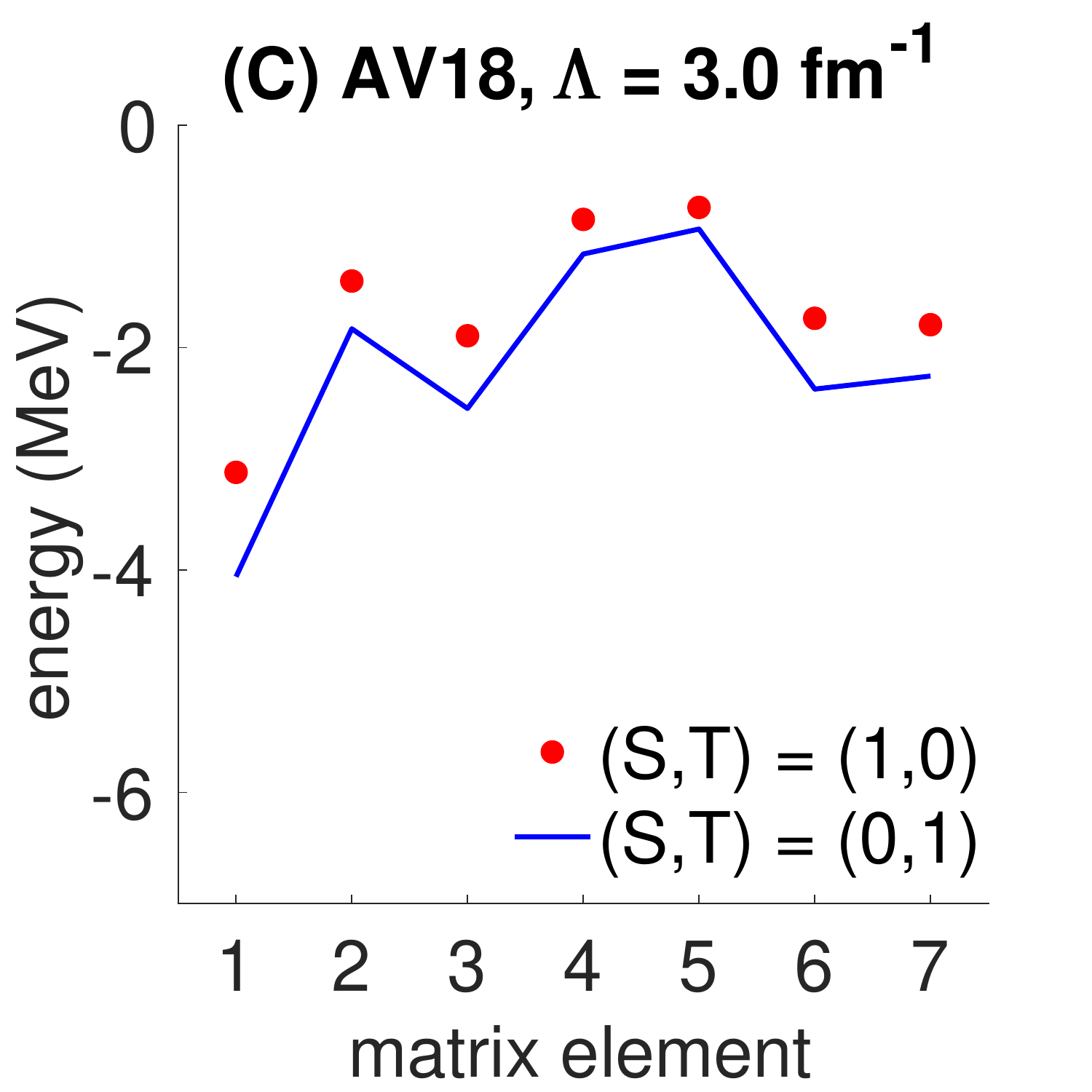}
\includegraphics[width=4cm]{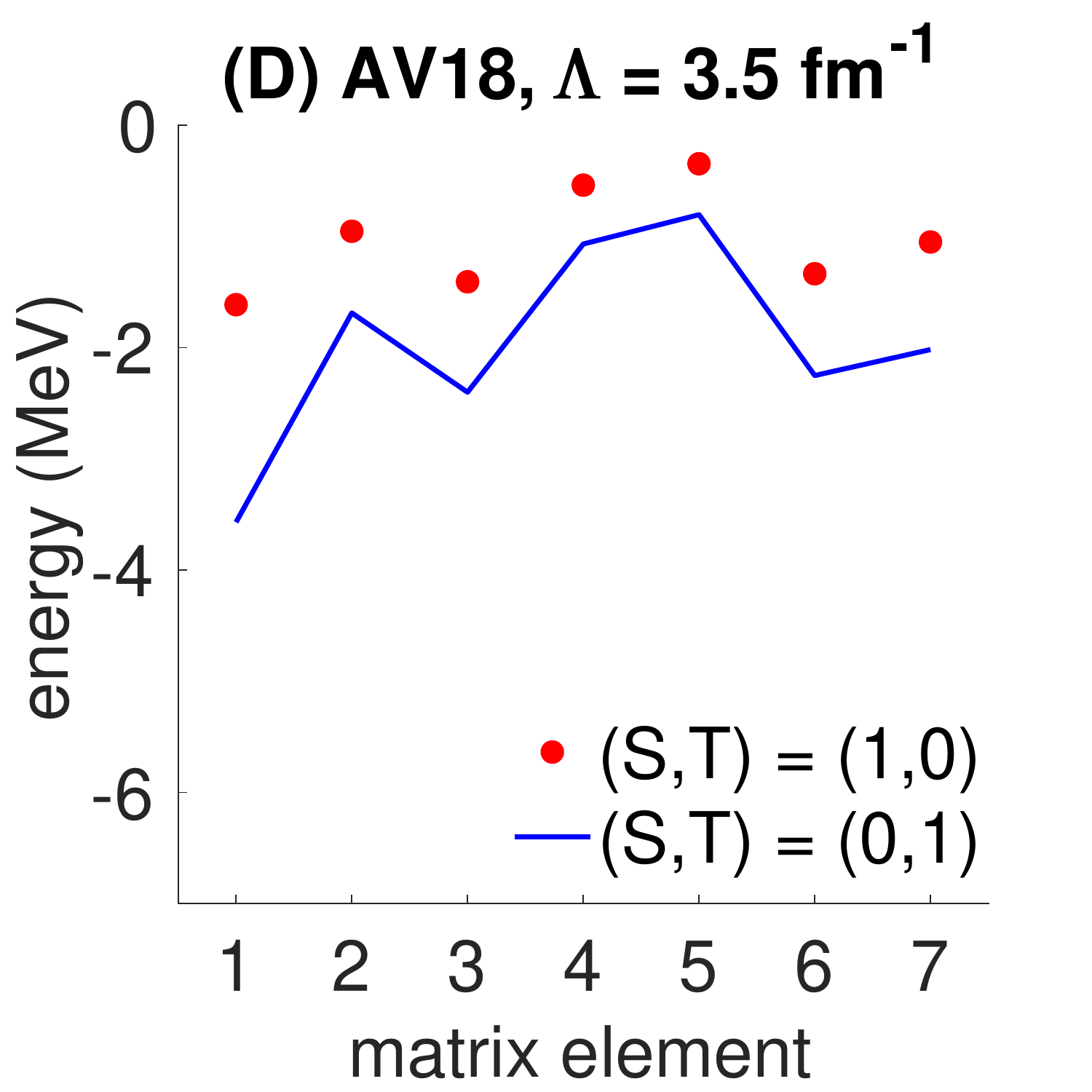}\\
\includegraphics[width=4cm]{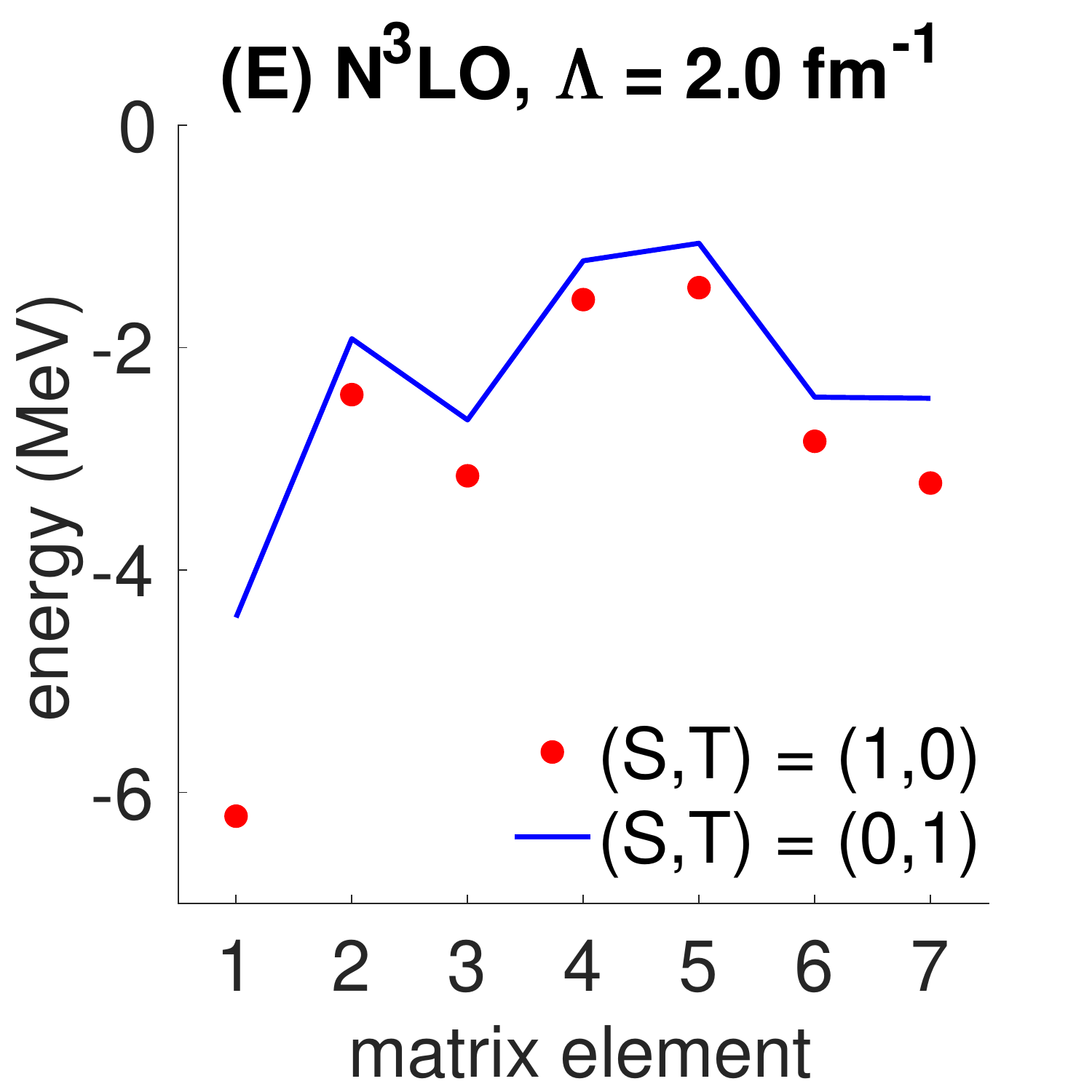}
\includegraphics[width=4cm]{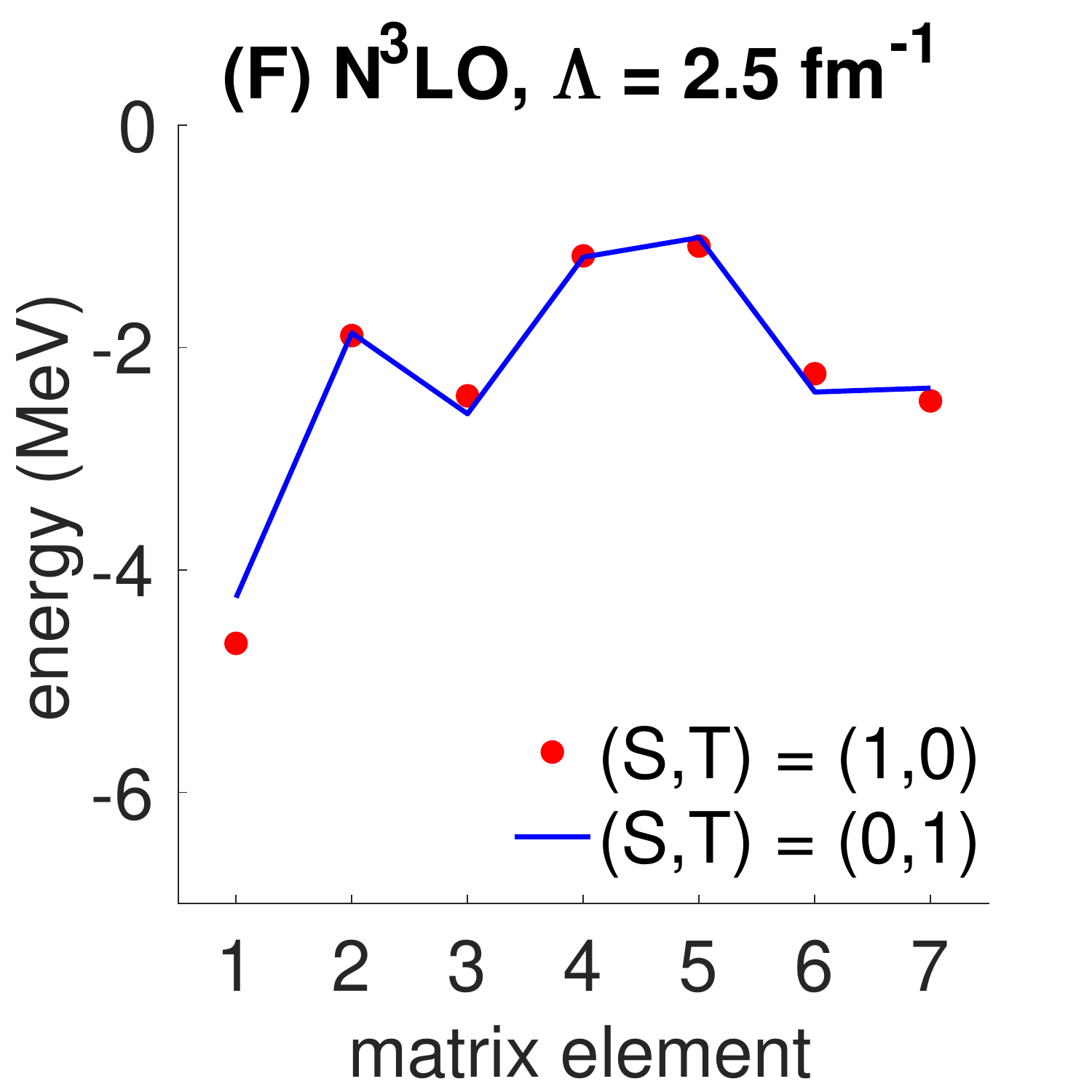}
\includegraphics[width=4cm]{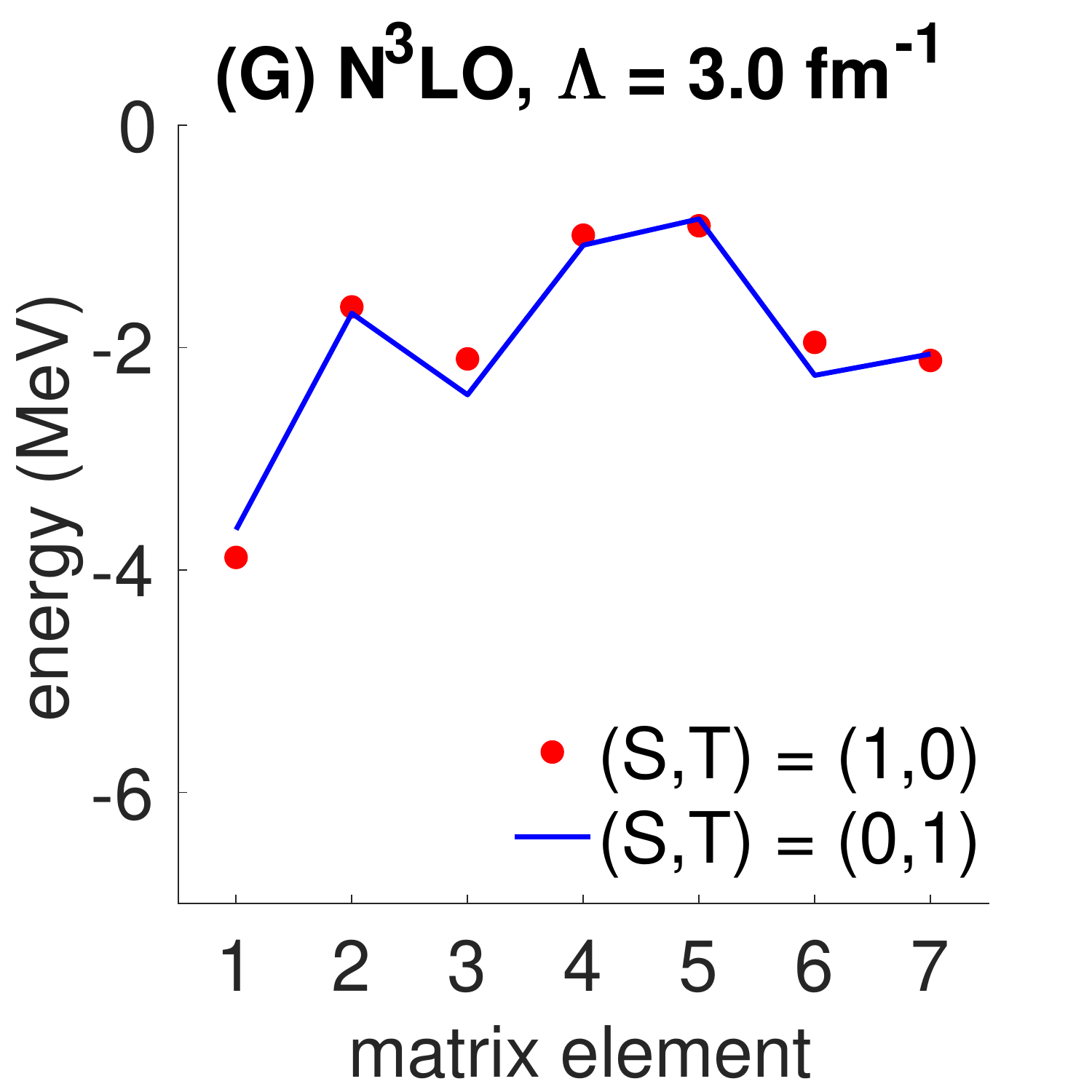}
\includegraphics[width=4cm]{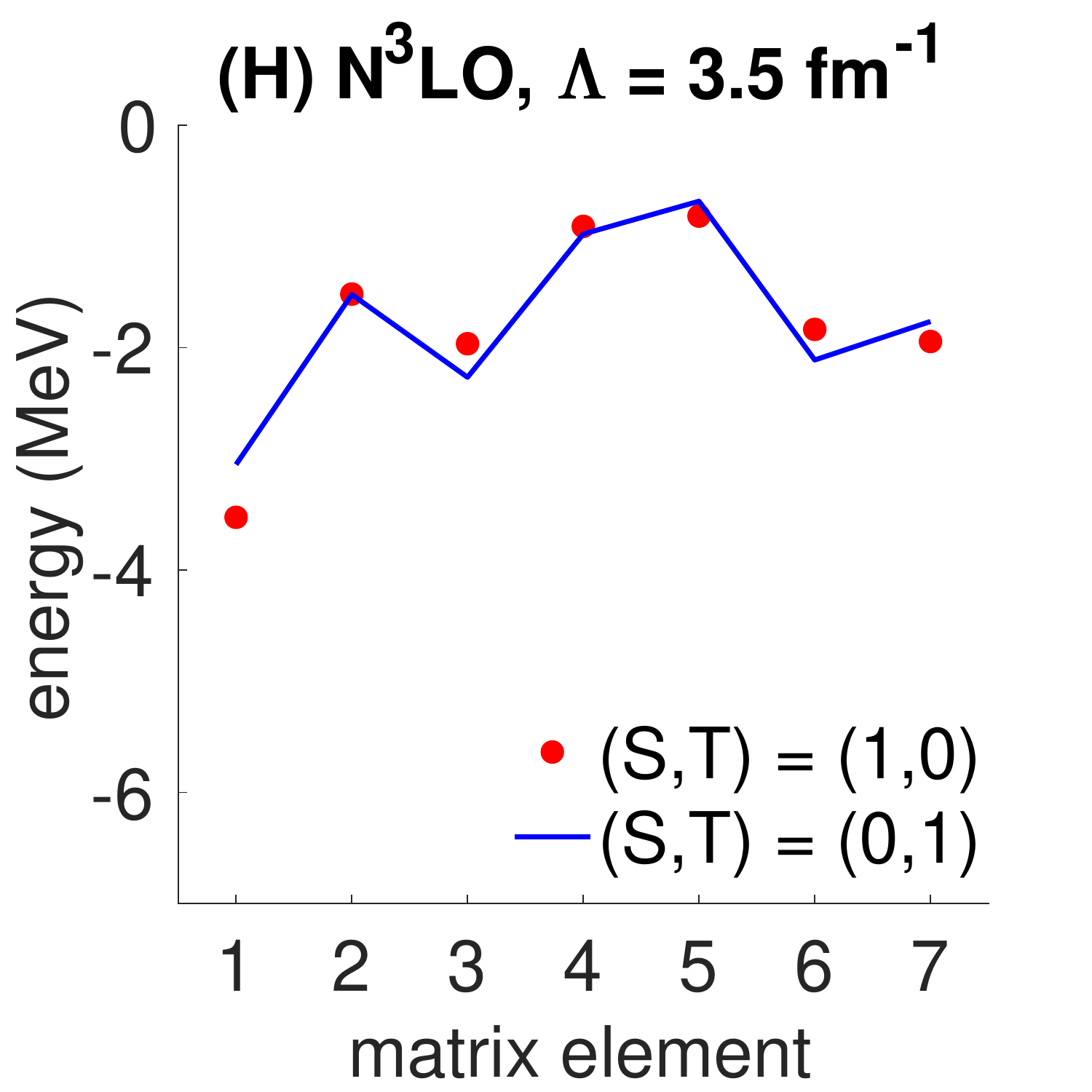}
\caption{(Color online) Two-body matrix elements for the 1s-0d shell. Panels (A-D):  Results for AV18 for $(S,T) = (1,0)$ with red dots and $(S,T) = (0,1)$ with blue lines at $\Lambda = 2.0~{\rm fm}^{-1}, 2.5~{\rm fm}^{-1}, 3.0~{\rm fm}^{-1}$, and $3.5~{\rm fm}^{-1}$, respectively.  Panels (E-H): Results for the chiral N$^3$LO interaction for $(S,T) = (1,0)$ with red dots and $(S,T) = (0,1)$ with blue lines at $\Lambda = 2.0~{\rm fm}^{-1}, 2.5~{\rm fm}^{-1}, 3.0~{\rm fm}^{-1}$, and $3.5~{\rm fm}^{-1}$, respectively.}
\label{TBME_2}
\end{figure*}

We can now test spin-isospin exchange for general two-body matrix elements in the 1s-0d shell.  For this analysis we use the spin-tensor analysis developed in Ref.~\cite{Kirson:1973ffz,Brown:1985,Brown:1988}.  We consider seven two-body matrix elements for $(S,T) = (1,0)$ and $(S,T) = (0,1)$.  The seven matrix elements are listed in Table~\ref{TBME_ST}.  $L_3$ and $L_4$ are the orbital angular momenta of the incoming orbitals corresponding to state $B$ in Eq.~\eqref{M_AB}, and $L_{34}$ is the total orbital angular momentum.  $L_1$ and $L_2$ are the orbital angular momenta of the outgoing orbitals corresponding to state $A$ in Eq.~\eqref{M_AB}, and $L_{12}$ is the total orbital angular momentum.  Matrix element $7$ is defined by the $1s_{1/2}$ orbitals that we have discussed above.  
\begin{table}
\caption{
Selected two-body matrix elements for the 1s-0d shell.\label{TBME_ST}
}
\centering{}%
\begin{tabular}{c | c c c c | c c }
\hline
matrix element & $L_1$ & $L_2$ & $L_3$ & $L_4$ & $L_{12}$ & $L_{34}$ \\
\hline
1 & 
2 & 2 & 2 & 2 & 0 & 0  \\
2 & 
2 & 2 & 2 & 2 & 2 & 2 \\
3 & 
2 & 2 & 2 & 2 & 4 & 4  \\
4 & 
2 & 2 & 2 & 0 & 2 & 2  \\
5 & 
2 & 2 & 0 & 0 & 0 & 0  \\
6 & 
2 & 0 & 2 & 0 & 2 & 2  \\
7 & 
0 & 0 & 0 & 0 & 0 & 0  \\
\hline
\end{tabular}
\end{table}
In addition to performing the average over $S_z$ in Eq.~\eqref{M_AB}, we also set $L_z=(L_{12})_z=(L_{34})_z$ and average over $L_z$ as well.  The results for AV18 and the chiral N$^3$LO interaction \cite{Entem:2003ft} at $\Lambda = 2.0, 2.5, 3.0, 3.5~{\rm fm}^{-1}$ are shown in Fig.~\ref{TBME_2}.  We see that the $(S,T)=(1,0)$ and $(S,T)=(0,1)$ results are nearly equal at $\Lambda = 2.5~{\rm fm}^{-1}$ for both AV18 and the N$^3$LO interaction, with a relative error of size $1/N_c^2$. We also confirm that for $\Lambda < \Lambda_{{\rm large-}N_c}$ the $(S,T) = (1,0)$ channel is more attractive, and for $\Lambda > \Lambda_{{\rm large-}N_c}$ the $(S,T) = (0,1)$ channel is more attractive.  We should point out that the chiral N$^3$LO interaction is defined with a momentum cutoff scale that is comparable to $\Lambda \sim 2.5~{\rm fm}^{-1}$, and so any renormalization scale dependence above this momentum value is relatively minor.  A detailed summary of the methods used in these shell model calculations can be found in Ref.~\cite{Brown:1988}, and a description of the $V_{{\rm low} k}$ effective shell model interactions can be found in Ref.~\cite{Bogner:1999my}.

The large-$N_c$ analysis can also be applied to three-nucleon interactions \cite{Phillips:2013rsa,Epelbaum:2014sea}.  In the large-$N_c$ limit, the leading central three-nucleon interactions will have the form
\begin{eqnarray}
V_{{\rm large-}N_c}^{\rm 3N} &=& V^{3{\rm  N}}_C + [(\vec{\sigma}_1\times \vec{\sigma}_2) \cdot \vec{\sigma}_3] [(\vec{\tau}_1 \times \vec{\tau}_2) \cdot \vec{\tau}_3] W^{3{\rm  N}}_{123} \nonumber \\
& +& \vec{\sigma}_1\cdot \vec{\sigma}_2 \vec{\tau}_1 \cdot \vec{\tau}_2 W^{3{\rm  N}}_{12}
+ \vec{\sigma}_2\cdot \vec{\sigma}_3 \vec{\tau}_2 \cdot \vec{\tau}_3 W^{3{\rm  N}}_{23}\nonumber  \\
& +& \vec{\sigma}_3\cdot \vec{\sigma}_1 \vec{\tau}_3 \cdot \vec{\tau}_1 W^{3{\rm  N}}_{31} + \dots, 
\label{3N}
\end{eqnarray}
provided that the momentum resolution scale is close to $\Lambda_{{\rm large-}N_c}$. In this case the subleading central three-nucleon interactions are of size $1/N_c$ due to terms such as $\vec{\sigma}_1\cdot \vec{\sigma}_2[(\vec{\tau}_1 \times \vec{\tau}_2) \cdot \vec{\tau}_3]$ and $[(\vec{\sigma}_1\times \vec{\sigma}_2) \cdot \vec{\sigma}_3]\vec{\tau}_1 \cdot \vec{\tau}_2$ \cite{Phillips:2013rsa}. This simplification should be helpful in constraining the many short-range three-nucleon interactions that appear at higher orders in chiral effective field theory.  The spin-isospin exchange symmetry of the leading interactions also severely limits the isospin-dependent contributions of the three-nucleon interactions to the nuclear equation of state.  This fact is relevant for calculations of the nuclear symmetry energy and its density dependence in dense nuclear matter.  

Spin-isospin exchange symmetry is also useful for constraining calculations that use local regularization to produce a nonzero interaction range for the three-nucleon contact interaction at N$^2$LO.  Local regularization produces local interactions with some strength in higher partial waves, and it has been observed that different isospin-dependent structures for the three-nucleon contact interaction produce different behaviors in many-body calculations \cite{Lonardoni:2017hgs,Lonardoni:2018nob}. The results of our analysis here suggest that such many-body calculations may be more reliable if one imposes spin-isospin exchange symmetry on the three-nucleon contact interaction for calculations with a momentum cutoff of $\Lambda_{{\rm large-}N_c}$.

In this work we have confirmed that the strong interactions have an approximate spin-isospin exchange symmetry that can be derived in the limit of many colors, $N_c$.  The derivation is valid with relative error $1/N_c^2$ only if the momentum resolution scale is near $\Lambda_{{\rm large-}N_c}\sim 500$~MeV.  While our work relies on the conceptual foundations provided by others, our findings may provide new insights into the empirical fact that chiral effective field theory is most effective for momentum cutoff scales near this value.  We believe that our findings provide new motivation for investigating and applying the constraints of the large-$N_c$ limit in future nuclear structure calculations. 

We have derived a set of spin-isospin exchange sum rules and discussed applications to the spectrum of $^{30}$P and as well as applications to nuclear forces, nuclear structure calculations, and constraining three-nucleon interactions.  Our results using modern forces and methods are nevertheless entirely consistent with the results presented in Fig.~3 of Ref.~\cite{Kaplan:1996rk}, which shows the dominance of meson exchange coupling terms in the Nijmegen $NN$ potential that respect spin-isospin exchange symmetry.
In the Supplemental Material we present evidence for spin-isospin exchange symmetry as seen from the nucleon-nucleon scattering phase shifts as well as further information about the lattice interactions.

\begin{acknowledgments}
We are grateful for illuminating discussions with Enrique Ruiz Arriola, David Kaplan, Witold Nazarewicz, Gautam Rupak, Martin Savage, and Robert Wiringa.  We acknowledge funding by the U.S. National Science Foundation (Grants No. PHY-1404159, PHY-1811855 and PHY-2013047), U.S. Department of Energy (DE-SC0017887 and DE-SC0018638), the Nuclear Computational Low-Energy
Initiative (NUCLEI) SciDAC-4 project (DE-SC0018083), 
Deutsche Forschungsgemeinschaft (Project-ID 196253076 - TRR 110),
Volkswagen\-Stiftung (93562),
the BMBF (Verbundprojekt 05P18PCFP1),
Chinese Academy of Sciences (CAS) through a President's International Fellowship Initiative (PIFI) (2018DM0034). The authors gratefully acknowledge computational resources provided by the Oak Ridge Leadership Computing
Facility through the INCITE award ``Ab-initio nuclear structure and nuclear reactions'', the Gauss Centre for Supercomputing e.V.
(www.gauss-centre.eu) for computing time on the GCS Supercomputer JUWELS at J{\"u}lich Supercomputing Centre (JSC), and Michigan
State University.
\end{acknowledgments}

\bibliography{References}

\begin{thebibliography}{51}
\expandafter\ifx\csname natexlab\endcsname\relax\def\natexlab#1{#1}\fi
\expandafter\ifx\csname bibnamefont\endcsname\relax
  \def\bibnamefont#1{#1}\fi
\expandafter\ifx\csname bibfnamefont\endcsname\relax
  \def\bibfnamefont#1{#1}\fi
\expandafter\ifx\csname citenamefont\endcsname\relax
  \def\citenamefont#1{#1}\fi
\expandafter\ifx\csname url\endcsname\relax
  \def\url#1{\texttt{#1}}\fi
\expandafter\ifx\csname urlprefix\endcsname\relax\def\urlprefix{URL }\fi
\providecommand{\bibinfo}[2]{#2}
\providecommand{\eprint}[2][]{\url{#2}}

\bibitem[{\citenamefont{'t~Hooft}(1974)}]{tHooft:1973alw}
\bibinfo{author}{\bibfnamefont{G.}~\bibnamefont{'t~Hooft}},
  \bibinfo{journal}{Nucl. Phys. B} \textbf{\bibinfo{volume}{72}},
  \bibinfo{pages}{461} (\bibinfo{year}{1974}).

\bibitem[{\citenamefont{Witten}(1979)}]{Witten:1979kh}
\bibinfo{author}{\bibfnamefont{E.}~\bibnamefont{Witten}},
  \bibinfo{journal}{Nucl. Phys. B} \textbf{\bibinfo{volume}{160}},
  \bibinfo{pages}{57} (\bibinfo{year}{1979}).

\bibitem[{\citenamefont{M{\"u}ther et~al.}(1987)\citenamefont{M{\"u}ther,
  Engelbrecht, and Brown}}]{Muther:1987sr}
\bibinfo{author}{\bibfnamefont{H.}~\bibnamefont{M{\"u}ther}},
  \bibinfo{author}{\bibfnamefont{C.}~\bibnamefont{Engelbrecht}},
  \bibnamefont{and} \bibinfo{author}{\bibfnamefont{G.}~\bibnamefont{Brown}},
  \bibinfo{journal}{Nucl. Phys. A} \textbf{\bibinfo{volume}{462}},
  \bibinfo{pages}{701} (\bibinfo{year}{1987}).

\bibitem[{\citenamefont{Dashen and Manohar}(1993)}]{Dashen:1993as}
\bibinfo{author}{\bibfnamefont{R.~F.} \bibnamefont{Dashen}} \bibnamefont{and}
  \bibinfo{author}{\bibfnamefont{A.~V.} \bibnamefont{Manohar}},
  \bibinfo{journal}{Phys. Lett. B} \textbf{\bibinfo{volume}{315}},
  \bibinfo{pages}{425} (\bibinfo{year}{1993}), \eprint{hep-ph/9307241}.

\bibitem[{\citenamefont{Dashen et~al.}(1995)\citenamefont{Dashen, Jenkins, and
  Manohar}}]{Dashen:1994qi}
\bibinfo{author}{\bibfnamefont{R.~F.} \bibnamefont{Dashen}},
  \bibinfo{author}{\bibfnamefont{E.~E.} \bibnamefont{Jenkins}},
  \bibnamefont{and} \bibinfo{author}{\bibfnamefont{A.~V.}
  \bibnamefont{Manohar}}, \bibinfo{journal}{Phys. Rev. D}
  \textbf{\bibinfo{volume}{51}}, \bibinfo{pages}{3697} (\bibinfo{year}{1995}),
  \eprint{hep-ph/9411234}.

\bibitem[{\citenamefont{Carone et~al.}(1994)\citenamefont{Carone, Georgi, and
  Osofsky}}]{Carone:1993dz}
\bibinfo{author}{\bibfnamefont{C.}~\bibnamefont{Carone}},
  \bibinfo{author}{\bibfnamefont{H.}~\bibnamefont{Georgi}}, \bibnamefont{and}
  \bibinfo{author}{\bibfnamefont{S.}~\bibnamefont{Osofsky}},
  \bibinfo{journal}{Phys. Lett. B} \textbf{\bibinfo{volume}{322}},
  \bibinfo{pages}{227} (\bibinfo{year}{1994}), \eprint{hep-ph/9310365}.

\bibitem[{\citenamefont{Jenkins and Manohar}(1994)}]{Jenkins:1994md}
\bibinfo{author}{\bibfnamefont{E.~E.} \bibnamefont{Jenkins}} \bibnamefont{and}
  \bibinfo{author}{\bibfnamefont{A.~V.} \bibnamefont{Manohar}},
  \bibinfo{journal}{Phys. Lett. B} \textbf{\bibinfo{volume}{335}},
  \bibinfo{pages}{452} (\bibinfo{year}{1994}), \eprint{hep-ph/9405431}.

\bibitem[{\citenamefont{Kaplan and Savage}(1996)}]{Kaplan:1995yg}
\bibinfo{author}{\bibfnamefont{D.~B.} \bibnamefont{Kaplan}} \bibnamefont{and}
  \bibinfo{author}{\bibfnamefont{M.~J.} \bibnamefont{Savage}},
  \bibinfo{journal}{Phys. Lett. B} \textbf{\bibinfo{volume}{365}},
  \bibinfo{pages}{244} (\bibinfo{year}{1996}), \eprint{hep-ph/9509371}.

\bibitem[{\citenamefont{Kaplan and Manohar}(1997)}]{Kaplan:1996rk}
\bibinfo{author}{\bibfnamefont{D.~B.} \bibnamefont{Kaplan}} \bibnamefont{and}
  \bibinfo{author}{\bibfnamefont{A.~V.} \bibnamefont{Manohar}},
  \bibinfo{journal}{Phys. Rev. C} \textbf{\bibinfo{volume}{56}},
  \bibinfo{pages}{76} (\bibinfo{year}{1997}), \eprint{nucl-th/9612021}.

\bibitem[{\citenamefont{Wigner}(1937)}]{Wigner:1937}
\bibinfo{author}{\bibfnamefont{E.}~\bibnamefont{Wigner}},
  \bibinfo{journal}{Phys. Rev.} \textbf{\bibinfo{volume}{51}},
  \bibinfo{pages}{106} (\bibinfo{year}{1937}).

\bibitem[{\citenamefont{Mehen et~al.}(1999)\citenamefont{Mehen, Stewart, and
  Wise}}]{Mehen:1999qs}
\bibinfo{author}{\bibfnamefont{T.}~\bibnamefont{Mehen}},
  \bibinfo{author}{\bibfnamefont{I.~W.} \bibnamefont{Stewart}},
  \bibnamefont{and} \bibinfo{author}{\bibfnamefont{M.~B.} \bibnamefont{Wise}},
  \bibinfo{journal}{Phys. Rev. Lett.} \textbf{\bibinfo{volume}{83}},
  \bibinfo{pages}{931} (\bibinfo{year}{1999}), \eprint{hep-ph/9902370}.

\bibitem[{\citenamefont{Chen et~al.}(2004)\citenamefont{Chen, Lee, and
  Sch{\"a}fer}}]{Chen:2004rq}
\bibinfo{author}{\bibfnamefont{J.-W.} \bibnamefont{Chen}},
  \bibinfo{author}{\bibfnamefont{D.}~\bibnamefont{Lee}}, \bibnamefont{and}
  \bibinfo{author}{\bibfnamefont{T.}~\bibnamefont{Sch{\"a}fer}},
  \bibinfo{journal}{Phys. Rev. Lett.} \textbf{\bibinfo{volume}{93}},
  \bibinfo{pages}{242302} (\bibinfo{year}{2004}), \eprint{nucl-th/0408043}.

\bibitem[{\citenamefont{Lee}(2007)}]{Lee:2007eu}
\bibinfo{author}{\bibfnamefont{D.}~\bibnamefont{Lee}}, \bibinfo{journal}{Phys.
  Rev. Lett.} \textbf{\bibinfo{volume}{98}}, \bibinfo{pages}{182501}
  (\bibinfo{year}{2007}), \eprint{nucl-th/0701041}.

\bibitem[{\citenamefont{Calle~Cordon and
  Ruiz~Arriola}(2009)}]{CalleCordon:2009ps}
\bibinfo{author}{\bibfnamefont{A.}~\bibnamefont{Calle~Cordon}}
  \bibnamefont{and}
  \bibinfo{author}{\bibfnamefont{E.}~\bibnamefont{Ruiz~Arriola}},
  \bibinfo{journal}{Phys. Rev. C} \textbf{\bibinfo{volume}{80}},
  \bibinfo{pages}{014002} (\bibinfo{year}{2009}), \eprint{0904.0421}.

\bibitem[{\citenamefont{K\"onig et~al.}(2017)\citenamefont{K\"onig,
  Grie\ss{}hammer, Hammer, and van Kolck}}]{Konig:2016utl}
\bibinfo{author}{\bibfnamefont{S.}~\bibnamefont{K\"onig}},
  \bibinfo{author}{\bibfnamefont{H.~W.} \bibnamefont{Grie\ss{}hammer}},
  \bibinfo{author}{\bibfnamefont{H.}~\bibnamefont{Hammer}}, \bibnamefont{and}
  \bibinfo{author}{\bibfnamefont{U.}~\bibnamefont{van Kolck}},
  \bibinfo{journal}{Phys. Rev. Lett.} \textbf{\bibinfo{volume}{118}},
  \bibinfo{pages}{202501} (\bibinfo{year}{2017}), \eprint{1607.04623}.

\bibitem[{\citenamefont{Lu et~al.}(2019)\citenamefont{Lu, Li, Elhatisari, Lee,
  Epelbaum, and Mei\ss{}ner}}]{Lu:2018bat}
\bibinfo{author}{\bibfnamefont{B.-N.} \bibnamefont{Lu}},
  \bibinfo{author}{\bibfnamefont{N.}~\bibnamefont{Li}},
  \bibinfo{author}{\bibfnamefont{S.}~\bibnamefont{Elhatisari}},
  \bibinfo{author}{\bibfnamefont{D.}~\bibnamefont{Lee}},
  \bibinfo{author}{\bibfnamefont{E.}~\bibnamefont{Epelbaum}}, \bibnamefont{and}
  \bibinfo{author}{\bibfnamefont{U.-G.} \bibnamefont{Mei\ss{}ner}},
  \bibinfo{journal}{Phys. Lett. B} \textbf{\bibinfo{volume}{797}},
  \bibinfo{pages}{134863} (\bibinfo{year}{2019}), \eprint{1812.10928}.

\bibitem[{\citenamefont{Karl and Paton}(1984)}]{Karl:1984cz}
\bibinfo{author}{\bibfnamefont{G.}~\bibnamefont{Karl}} \bibnamefont{and}
  \bibinfo{author}{\bibfnamefont{J.~E.} \bibnamefont{Paton}},
  \bibinfo{journal}{Phys. Rev. D} \textbf{\bibinfo{volume}{30}},
  \bibinfo{pages}{238} (\bibinfo{year}{1984}).

\bibitem[{\citenamefont{Mergell et~al.}(1996)\citenamefont{Mergell,
  Mei{\ss}ner, and Drechsel}}]{Mergell:1995bf}
\bibinfo{author}{\bibfnamefont{P.}~\bibnamefont{Mergell}},
  \bibinfo{author}{\bibfnamefont{U.-G.} \bibnamefont{Mei{\ss}ner}},
  \bibnamefont{and} \bibinfo{author}{\bibfnamefont{D.}~\bibnamefont{Drechsel}},
  \bibinfo{journal}{Nucl. Phys. A} \textbf{\bibinfo{volume}{596}},
  \bibinfo{pages}{367} (\bibinfo{year}{1996}), \eprint{hep-ph/9506375}.

\bibitem[{\citenamefont{Belushkin et~al.}(2007)\citenamefont{Belushkin, Hammer,
  and Mei{\ss}ner}}]{Belushkin:2006qa}
\bibinfo{author}{\bibfnamefont{M.}~\bibnamefont{Belushkin}},
  \bibinfo{author}{\bibfnamefont{H.-W.} \bibnamefont{Hammer}},
  \bibnamefont{and} \bibinfo{author}{\bibfnamefont{U.-G.}
  \bibnamefont{Mei{\ss}ner}}, \bibinfo{journal}{Phys. Rev. C}
  \textbf{\bibinfo{volume}{75}}, \bibinfo{pages}{035202}
  (\bibinfo{year}{2007}), \eprint{hep-ph/0608337}.

\bibitem[{\citenamefont{Shanahan et~al.}(2014)\citenamefont{Shanahan, Thomas,
  Young, Zanotti, Horsley, Nakamura, Pleiter, Rakow, Schierholz, and
  St\"uben}}]{Shanahan:2014uka}
\bibinfo{author}{\bibfnamefont{P.}~\bibnamefont{Shanahan}},
  \bibinfo{author}{\bibfnamefont{A.}~\bibnamefont{Thomas}},
  \bibinfo{author}{\bibfnamefont{R.}~\bibnamefont{Young}},
  \bibinfo{author}{\bibfnamefont{J.}~\bibnamefont{Zanotti}},
  \bibinfo{author}{\bibfnamefont{R.}~\bibnamefont{Horsley}},
  \bibinfo{author}{\bibfnamefont{Y.}~\bibnamefont{Nakamura}},
  \bibinfo{author}{\bibfnamefont{D.}~\bibnamefont{Pleiter}},
  \bibinfo{author}{\bibfnamefont{P.}~\bibnamefont{Rakow}},
  \bibinfo{author}{\bibfnamefont{G.}~\bibnamefont{Schierholz}},
  \bibnamefont{and} \bibinfo{author}{\bibfnamefont{H.}~\bibnamefont{St\"uben}}
  (\bibinfo{collaboration}{CSSM, QCDSF/UKQCD}), \bibinfo{journal}{Phys. Rev. D}
  \textbf{\bibinfo{volume}{89}}, \bibinfo{pages}{074511}
  (\bibinfo{year}{2014}), \eprint{1401.5862}.

\bibitem[{\citenamefont{Parreno et~al.}(2017)\citenamefont{Parreno, Savage,
  Tiburzi, Wilhelm, Chang, Detmold, and Orginos}}]{Parreno:2016fwu}
\bibinfo{author}{\bibfnamefont{A.}~\bibnamefont{Parreno}},
  \bibinfo{author}{\bibfnamefont{M.~J.} \bibnamefont{Savage}},
  \bibinfo{author}{\bibfnamefont{B.~C.} \bibnamefont{Tiburzi}},
  \bibinfo{author}{\bibfnamefont{J.}~\bibnamefont{Wilhelm}},
  \bibinfo{author}{\bibfnamefont{E.}~\bibnamefont{Chang}},
  \bibinfo{author}{\bibfnamefont{W.}~\bibnamefont{Detmold}}, \bibnamefont{and}
  \bibinfo{author}{\bibfnamefont{K.}~\bibnamefont{Orginos}},
  \bibinfo{journal}{Phys. Rev. D} \textbf{\bibinfo{volume}{95}},
  \bibinfo{pages}{114513} (\bibinfo{year}{2017}), \eprint{1609.03985}.

\bibitem[{\citenamefont{Wagman et~al.}(2017)\citenamefont{Wagman, Winter,
  Chang, Davoudi, Detmold, Orginos, Savage, and Shanahan}}]{Wagman:2017tmp}
\bibinfo{author}{\bibfnamefont{M.~L.} \bibnamefont{Wagman}},
  \bibinfo{author}{\bibfnamefont{F.}~\bibnamefont{Winter}},
  \bibinfo{author}{\bibfnamefont{E.}~\bibnamefont{Chang}},
  \bibinfo{author}{\bibfnamefont{Z.}~\bibnamefont{Davoudi}},
  \bibinfo{author}{\bibfnamefont{W.}~\bibnamefont{Detmold}},
  \bibinfo{author}{\bibfnamefont{K.}~\bibnamefont{Orginos}},
  \bibinfo{author}{\bibfnamefont{M.~J.} \bibnamefont{Savage}},
  \bibnamefont{and} \bibinfo{author}{\bibfnamefont{P.~E.}
  \bibnamefont{Shanahan}}, \bibinfo{journal}{Phys. Rev. D}
  \textbf{\bibinfo{volume}{96}}, \bibinfo{pages}{114510}
  (\bibinfo{year}{2017}), \eprint{1706.06550}.

\bibitem[{\citenamefont{Illa et~al.}(2020)}]{Illa:2020nsi}
\bibinfo{author}{\bibfnamefont{M.}~\bibnamefont{Illa}} \bibnamefont{et~al.}
  (\bibinfo{year}{2020}), \eprint{2009.12357}.

\bibitem[{\citenamefont{Banerjee et~al.}(2002)\citenamefont{Banerjee, Cohen,
  and Gelman}}]{Banerjee:2001js}
\bibinfo{author}{\bibfnamefont{M.~K.} \bibnamefont{Banerjee}},
  \bibinfo{author}{\bibfnamefont{T.~D.} \bibnamefont{Cohen}}, \bibnamefont{and}
  \bibinfo{author}{\bibfnamefont{B.~A.} \bibnamefont{Gelman}},
  \bibinfo{journal}{Phys. Rev. C} \textbf{\bibinfo{volume}{65}},
  \bibinfo{pages}{034011} (\bibinfo{year}{2002}), \eprint{hep-ph/0109274}.

\bibitem[{\citenamefont{Cohen}(2002)}]{Cohen:2002im}
\bibinfo{author}{\bibfnamefont{T.~D.} \bibnamefont{Cohen}},
  \bibinfo{journal}{Phys. Rev. C} \textbf{\bibinfo{volume}{66}},
  \bibinfo{pages}{064003} (\bibinfo{year}{2002}), \eprint{nucl-th/0209072}.

\bibitem[{\citenamefont{Cohen and Gelman}(2002)}]{Cohen:2002qn}
\bibinfo{author}{\bibfnamefont{T.~D.} \bibnamefont{Cohen}} \bibnamefont{and}
  \bibinfo{author}{\bibfnamefont{B.~A.} \bibnamefont{Gelman}},
  \bibinfo{journal}{Phys. Lett. B} \textbf{\bibinfo{volume}{540}},
  \bibinfo{pages}{227} (\bibinfo{year}{2002}), \eprint{nucl-th/0202036}.

\bibitem[{\citenamefont{Kaplan et~al.}(1998{\natexlab{a}})\citenamefont{Kaplan,
  Savage, and Wise}}]{Kaplan:1998tg}
\bibinfo{author}{\bibfnamefont{D.~B.} \bibnamefont{Kaplan}},
  \bibinfo{author}{\bibfnamefont{M.~J.} \bibnamefont{Savage}},
  \bibnamefont{and} \bibinfo{author}{\bibfnamefont{M.~B.} \bibnamefont{Wise}},
  \bibinfo{journal}{Phys. Lett. B} \textbf{\bibinfo{volume}{424}},
  \bibinfo{pages}{390} (\bibinfo{year}{1998}{\natexlab{a}}),
  \eprint{nucl-th/9801034}.

\bibitem[{\citenamefont{Kaplan et~al.}(1998{\natexlab{b}})\citenamefont{Kaplan,
  Savage, and Wise}}]{Kaplan:1998we}
\bibinfo{author}{\bibfnamefont{D.~B.} \bibnamefont{Kaplan}},
  \bibinfo{author}{\bibfnamefont{M.~J.} \bibnamefont{Savage}},
  \bibnamefont{and} \bibinfo{author}{\bibfnamefont{M.~B.} \bibnamefont{Wise}},
  \bibinfo{journal}{Nucl. Phys. B} \textbf{\bibinfo{volume}{534}},
  \bibinfo{pages}{329} (\bibinfo{year}{1998}{\natexlab{b}}),
  \eprint{nucl-th/9802075}.

\bibitem[{\citenamefont{Timoteo et~al.}(2012)\citenamefont{Timoteo, Szpigel,
  and Ruiz~Arriola}}]{Timoteo:2011tt}
\bibinfo{author}{\bibfnamefont{V.}~\bibnamefont{Timoteo}},
  \bibinfo{author}{\bibfnamefont{S.}~\bibnamefont{Szpigel}}, \bibnamefont{and}
  \bibinfo{author}{\bibfnamefont{E.}~\bibnamefont{Ruiz~Arriola}},
  \bibinfo{journal}{Phys. Rev. C} \textbf{\bibinfo{volume}{86}},
  \bibinfo{pages}{034002} (\bibinfo{year}{2012}), \eprint{1108.1162}.

\bibitem[{\citenamefont{Ruiz~Arriola et~al.}(2013)\citenamefont{Ruiz~Arriola,
  Timoteo, and Szpigel}}]{Arriola:2013nja}
\bibinfo{author}{\bibfnamefont{E.}~\bibnamefont{Ruiz~Arriola}},
  \bibinfo{author}{\bibfnamefont{V.}~\bibnamefont{Timoteo}}, \bibnamefont{and}
  \bibinfo{author}{\bibfnamefont{S.}~\bibnamefont{Szpigel}},
  \bibinfo{journal}{PoS} \textbf{\bibinfo{volume}{CD12}}, \bibinfo{pages}{106}
  (\bibinfo{year}{2013}), \eprint{1302.3978}.

\bibitem[{\citenamefont{Ruiz~Arriola}(2016)}]{RuizArriola:2016vap}
\bibinfo{author}{\bibfnamefont{E.}~\bibnamefont{Ruiz~Arriola}},
  \bibinfo{journal}{Symmetry} \textbf{\bibinfo{volume}{8}}, \bibinfo{pages}{42}
  (\bibinfo{year}{2016}).

\bibitem[{\citenamefont{Schindler et~al.}(2018)\citenamefont{Schindler, Singh,
  and Springer}}]{Schindler:2018irz}
\bibinfo{author}{\bibfnamefont{M.~R.} \bibnamefont{Schindler}},
  \bibinfo{author}{\bibfnamefont{H.}~\bibnamefont{Singh}}, \bibnamefont{and}
  \bibinfo{author}{\bibfnamefont{R.~P.} \bibnamefont{Springer}},
  \bibinfo{journal}{Phys. Rev. C} \textbf{\bibinfo{volume}{98}},
  \bibinfo{pages}{044001} (\bibinfo{year}{2018}), \eprint{1805.06056}.

\bibitem[{\citenamefont{Beane et~al.}(2019)\citenamefont{Beane, Kaplan, Klco,
  and Savage}}]{Beane:2018oxh}
\bibinfo{author}{\bibfnamefont{S.~R.} \bibnamefont{Beane}},
  \bibinfo{author}{\bibfnamefont{D.~B.} \bibnamefont{Kaplan}},
  \bibinfo{author}{\bibfnamefont{N.}~\bibnamefont{Klco}}, \bibnamefont{and}
  \bibinfo{author}{\bibfnamefont{M.~J.} \bibnamefont{Savage}},
  \bibinfo{journal}{Phys. Rev. Lett.} \textbf{\bibinfo{volume}{122}},
  \bibinfo{pages}{102001} (\bibinfo{year}{2019}), \eprint{1812.03138}.

\bibitem[{\citenamefont{Gezerlis et~al.}(2014)\citenamefont{Gezerlis, Tews,
  Epelbaum, Freunek, Gandolfi, Hebeler, Nogga, and Schwenk}}]{Gezerlis:2014zia}
\bibinfo{author}{\bibfnamefont{A.}~\bibnamefont{Gezerlis}},
  \bibinfo{author}{\bibfnamefont{I.}~\bibnamefont{Tews}},
  \bibinfo{author}{\bibfnamefont{E.}~\bibnamefont{Epelbaum}},
  \bibinfo{author}{\bibfnamefont{M.}~\bibnamefont{Freunek}},
  \bibinfo{author}{\bibfnamefont{S.}~\bibnamefont{Gandolfi}},
  \bibinfo{author}{\bibfnamefont{K.}~\bibnamefont{Hebeler}},
  \bibinfo{author}{\bibfnamefont{A.}~\bibnamefont{Nogga}}, \bibnamefont{and}
  \bibinfo{author}{\bibfnamefont{A.}~\bibnamefont{Schwenk}},
  \bibinfo{journal}{Phys. Rev. C} \textbf{\bibinfo{volume}{90}},
  \bibinfo{pages}{054323} (\bibinfo{year}{2014}), \eprint{1406.0454}.

\bibitem[{\citenamefont{Reinert et~al.}(2018)\citenamefont{Reinert, Krebs, and
  Epelbaum}}]{Reinert:2017usi}
\bibinfo{author}{\bibfnamefont{P.}~\bibnamefont{Reinert}},
  \bibinfo{author}{\bibfnamefont{H.}~\bibnamefont{Krebs}}, \bibnamefont{and}
  \bibinfo{author}{\bibfnamefont{E.}~\bibnamefont{Epelbaum}},
  \bibinfo{journal}{Eur. Phys. J. A} \textbf{\bibinfo{volume}{54}},
  \bibinfo{pages}{86} (\bibinfo{year}{2018}), \eprint{1711.08821}.

\bibitem[{\citenamefont{Wiringa et~al.}(1995)\citenamefont{Wiringa, Stoks, and
  Schiavilla}}]{Wiringa:1994wb}
\bibinfo{author}{\bibfnamefont{R.~B.} \bibnamefont{Wiringa}},
  \bibinfo{author}{\bibfnamefont{V.}~\bibnamefont{Stoks}}, \bibnamefont{and}
  \bibinfo{author}{\bibfnamefont{R.}~\bibnamefont{Schiavilla}},
  \bibinfo{journal}{Phys. Rev. C} \textbf{\bibinfo{volume}{51}},
  \bibinfo{pages}{38} (\bibinfo{year}{1995}), \eprint{nucl-th/9408016}.

\bibitem[{\citenamefont{Bogner et~al.}(2002)\citenamefont{Bogner, Kuo,
  Coraggio, Covello, and Itaco}}]{Bogner:1999my}
\bibinfo{author}{\bibfnamefont{S.}~\bibnamefont{Bogner}},
  \bibinfo{author}{\bibfnamefont{T.}~\bibnamefont{Kuo}},
  \bibinfo{author}{\bibfnamefont{L.}~\bibnamefont{Coraggio}},
  \bibinfo{author}{\bibfnamefont{A.}~\bibnamefont{Covello}}, \bibnamefont{and}
  \bibinfo{author}{\bibfnamefont{N.}~\bibnamefont{Itaco}},
  \bibinfo{journal}{Phys. Rev. C} \textbf{\bibinfo{volume}{65}},
  \bibinfo{pages}{051301} (\bibinfo{year}{2002}), \eprint{nucl-th/9912056}.

\bibitem[{\citenamefont{Entem and Machleidt}(2003)}]{Entem:2003ft}
\bibinfo{author}{\bibfnamefont{D.}~\bibnamefont{Entem}} \bibnamefont{and}
  \bibinfo{author}{\bibfnamefont{R.}~\bibnamefont{Machleidt}},
  \bibinfo{journal}{Phys. Rev. C} \textbf{\bibinfo{volume}{68}},
  \bibinfo{pages}{041001} (\bibinfo{year}{2003}), \eprint{nucl-th/0304018}.

\bibitem[{\citenamefont{Kirson}(1973)}]{Kirson:1973ffz}
\bibinfo{author}{\bibfnamefont{M.}~\bibnamefont{Kirson}},
  \bibinfo{journal}{Phys. Lett. B} \textbf{\bibinfo{volume}{47}},
  \bibinfo{pages}{110} (\bibinfo{year}{1973}).

\bibitem[{\citenamefont{Brown et~al.}(1985)\citenamefont{Brown, Richter, and
  Wildenthal}}]{Brown:1985}
\bibinfo{author}{\bibfnamefont{B.~A.} \bibnamefont{Brown}},
  \bibinfo{author}{\bibfnamefont{W.~A.} \bibnamefont{Richter}},
  \bibnamefont{and} \bibinfo{author}{\bibfnamefont{B.~H.}
  \bibnamefont{Wildenthal}}, \bibinfo{journal}{Journal of Physics G: Nuclear
  Physics} \textbf{\bibinfo{volume}{11}}, \bibinfo{pages}{1191}
  (\bibinfo{year}{1985}).

\bibitem[{\citenamefont{Brown et~al.}(1988)\citenamefont{Brown, Richter,
  Julies, and Wildenthal}}]{Brown:1988}
\bibinfo{author}{\bibfnamefont{B.~A.} \bibnamefont{Brown}},
  \bibinfo{author}{\bibfnamefont{W.~A.} \bibnamefont{Richter}},
  \bibinfo{author}{\bibfnamefont{R.~E.} \bibnamefont{Julies}},
  \bibnamefont{and}
  \bibinfo{author}{\bibfnamefont{B.}~\bibnamefont{Wildenthal}},
  \bibinfo{journal}{Annals of Physics} \textbf{\bibinfo{volume}{182}},
  \bibinfo{pages}{191 } (\bibinfo{year}{1988}), ISSN \bibinfo{issn}{0003-4916}.

\bibitem[{\citenamefont{Phillips and Schat}(2013)}]{Phillips:2013rsa}
\bibinfo{author}{\bibfnamefont{D.~R.} \bibnamefont{Phillips}} \bibnamefont{and}
  \bibinfo{author}{\bibfnamefont{C.}~\bibnamefont{Schat}},
  \bibinfo{journal}{Phys. Rev. C} \textbf{\bibinfo{volume}{88}},
  \bibinfo{pages}{034002} (\bibinfo{year}{2013}), \eprint{1307.6274}.

\bibitem[{\citenamefont{Epelbaum et~al.}(2015)\citenamefont{Epelbaum,
  Gasparyan, Krebs, and Schat}}]{Epelbaum:2014sea}
\bibinfo{author}{\bibfnamefont{E.}~\bibnamefont{Epelbaum}},
  \bibinfo{author}{\bibfnamefont{A.}~\bibnamefont{Gasparyan}},
  \bibinfo{author}{\bibfnamefont{H.}~\bibnamefont{Krebs}}, \bibnamefont{and}
  \bibinfo{author}{\bibfnamefont{C.}~\bibnamefont{Schat}},
  \bibinfo{journal}{Eur. Phys. J. A} \textbf{\bibinfo{volume}{51}},
  \bibinfo{pages}{26} (\bibinfo{year}{2015}), \eprint{1411.3612}.

\bibitem[{\citenamefont{Lonardoni
  et~al.}(2018{\natexlab{a}})\citenamefont{Lonardoni, Carlson, Gandolfi, Lynn,
  Schmidt, Schwenk, and Wang}}]{Lonardoni:2017hgs}
\bibinfo{author}{\bibfnamefont{D.}~\bibnamefont{Lonardoni}},
  \bibinfo{author}{\bibfnamefont{J.}~\bibnamefont{Carlson}},
  \bibinfo{author}{\bibfnamefont{S.}~\bibnamefont{Gandolfi}},
  \bibinfo{author}{\bibfnamefont{J.}~\bibnamefont{Lynn}},
  \bibinfo{author}{\bibfnamefont{K.}~\bibnamefont{Schmidt}},
  \bibinfo{author}{\bibfnamefont{A.}~\bibnamefont{Schwenk}}, \bibnamefont{and}
  \bibinfo{author}{\bibfnamefont{X.}~\bibnamefont{Wang}},
  \bibinfo{journal}{Phys. Rev. Lett.} \textbf{\bibinfo{volume}{120}},
  \bibinfo{pages}{122502} (\bibinfo{year}{2018}{\natexlab{a}}),
  \eprint{1709.09143}.

\bibitem[{\citenamefont{Lonardoni
  et~al.}(2018{\natexlab{b}})\citenamefont{Lonardoni, Gandolfi, Lynn, Petrie,
  Carlson, Schmidt, and Schwenk}}]{Lonardoni:2018nob}
\bibinfo{author}{\bibfnamefont{D.}~\bibnamefont{Lonardoni}},
  \bibinfo{author}{\bibfnamefont{S.}~\bibnamefont{Gandolfi}},
  \bibinfo{author}{\bibfnamefont{J.}~\bibnamefont{Lynn}},
  \bibinfo{author}{\bibfnamefont{C.}~\bibnamefont{Petrie}},
  \bibinfo{author}{\bibfnamefont{J.}~\bibnamefont{Carlson}},
  \bibinfo{author}{\bibfnamefont{K.}~\bibnamefont{Schmidt}}, \bibnamefont{and}
  \bibinfo{author}{\bibfnamefont{A.}~\bibnamefont{Schwenk}},
  \bibinfo{journal}{Phys. Rev. C} \textbf{\bibinfo{volume}{97}},
  \bibinfo{pages}{044318} (\bibinfo{year}{2018}{\natexlab{b}}),
  \eprint{1802.08932}.

\bibitem[{\citenamefont{Stoks et~al.}(1993)\citenamefont{Stoks, Klomp,
  Rentmeester, and de~Swart}}]{Stoks:1993tb}
\bibinfo{author}{\bibfnamefont{V.}~\bibnamefont{Stoks}},
  \bibinfo{author}{\bibfnamefont{R.}~\bibnamefont{Klomp}},
  \bibinfo{author}{\bibfnamefont{M.}~\bibnamefont{Rentmeester}},
  \bibnamefont{and} \bibinfo{author}{\bibfnamefont{J.}~\bibnamefont{de~Swart}},
  \bibinfo{journal}{Phys. Rev. C} \textbf{\bibinfo{volume}{48}},
  \bibinfo{pages}{792} (\bibinfo{year}{1993}).

\bibitem[{\citenamefont{Machleidt}(2001)}]{Machleidt:2000ge}
\bibinfo{author}{\bibfnamefont{R.}~\bibnamefont{Machleidt}},
  \bibinfo{journal}{Phys. Rev. C} \textbf{\bibinfo{volume}{63}},
  \bibinfo{pages}{024001} (\bibinfo{year}{2001}), \eprint{nucl-th/0006014}.

\bibitem[{\citenamefont{Entem et~al.}(2017)\citenamefont{Entem, Machleidt, and
  Nosyk}}]{Entem:2017gor}
\bibinfo{author}{\bibfnamefont{D.}~\bibnamefont{Entem}},
  \bibinfo{author}{\bibfnamefont{R.}~\bibnamefont{Machleidt}},
  \bibnamefont{and} \bibinfo{author}{\bibfnamefont{Y.}~\bibnamefont{Nosyk}},
  \bibinfo{journal}{Phys. Rev. C} \textbf{\bibinfo{volume}{96}},
  \bibinfo{pages}{024004} (\bibinfo{year}{2017}), \eprint{1703.05454}.

\bibitem[{\citenamefont{Li et~al.}(2018)\citenamefont{Li, Elhatisari, Epelbaum,
  Lee, Lu, and Mei{\ss}ner}}]{Li:2018ymw}
\bibinfo{author}{\bibfnamefont{N.}~\bibnamefont{Li}},
  \bibinfo{author}{\bibfnamefont{S.}~\bibnamefont{Elhatisari}},
  \bibinfo{author}{\bibfnamefont{E.}~\bibnamefont{Epelbaum}},
  \bibinfo{author}{\bibfnamefont{D.}~\bibnamefont{Lee}},
  \bibinfo{author}{\bibfnamefont{B.-N.} \bibnamefont{Lu}}, \bibnamefont{and}
  \bibinfo{author}{\bibfnamefont{U.-G.} \bibnamefont{Mei{\ss}ner}},
  \bibinfo{journal}{Phys. Rev. C} \textbf{\bibinfo{volume}{98}},
  \bibinfo{pages}{044002} (\bibinfo{year}{2018}), \eprint{1806.07994}.

\bibitem[{\citenamefont{Lee}(2009)}]{Lee:2008fa}
\bibinfo{author}{\bibfnamefont{D.}~\bibnamefont{Lee}}, \bibinfo{journal}{Prog.
  Part. Nucl. Phys.} \textbf{\bibinfo{volume}{63}}, \bibinfo{pages}{117}
  (\bibinfo{year}{2009}), \eprint{0804.3501}.

\bibitem[{\citenamefont{L\"ahde and Mei\ss{}ner}(2019)}]{Lahde:2019npb}
\bibinfo{author}{\bibfnamefont{T.~A.} \bibnamefont{L\"ahde}} \bibnamefont{and}
  \bibinfo{author}{\bibfnamefont{U.-G.} \bibnamefont{Mei\ss{}ner}},
  \emph{\bibinfo{title}{{Nuclear Lattice Effective Field Theory}: {An
  introduction}}}, vol. \bibinfo{volume}{957} (\bibinfo{publisher}{Springer},
  \bibinfo{year}{2019}), ISBN \bibinfo{isbn}{978-3-030-14187-5,
  978-3-030-14189-9}.

\end{thebibliography}
\bibliographystyle{apsrev}

\clearpage    

\beginsupplement

\onecolumngrid

\section{Supplemental Material}
\subsection{Phase shifts}
\begin{figure}[htb]
                \centering
 \caption{S-wave lattice phase shifts for neutron-proton scattering.  The left panel shows the spin-singlet channel.  The  right panel shows results for the spin-triplet channel and the spin-triplet results when restricted to the $L=0$ orbital angular momentum channel.}
\includegraphics[width=12.0cm]{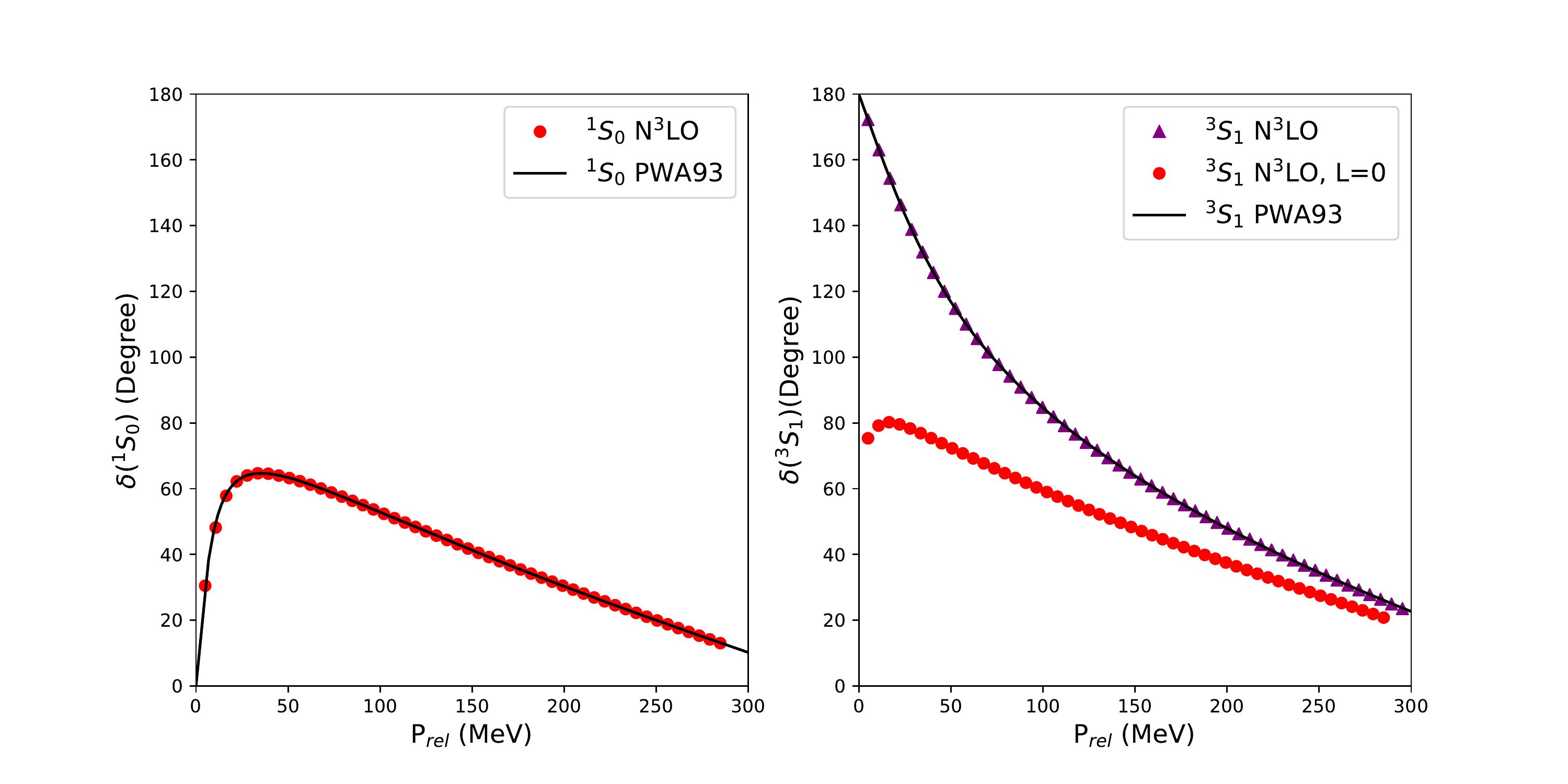}
                \label{S-wave}
        \end{figure}

\begin{figure}[!h]
                \centering
\caption{D-wave lattice phase shifts for neutron-proton scattering.  The left panel shows the spin-singlet channel.  The  right panel shows the spin-triplet channels and the spin-triplet results when restricted to the $L=2$ orbital angular momentum channel and averaged over all possible values for the total angular momentum quantum numbers $J,J_z$. }                 
\includegraphics[width=12.0cm]{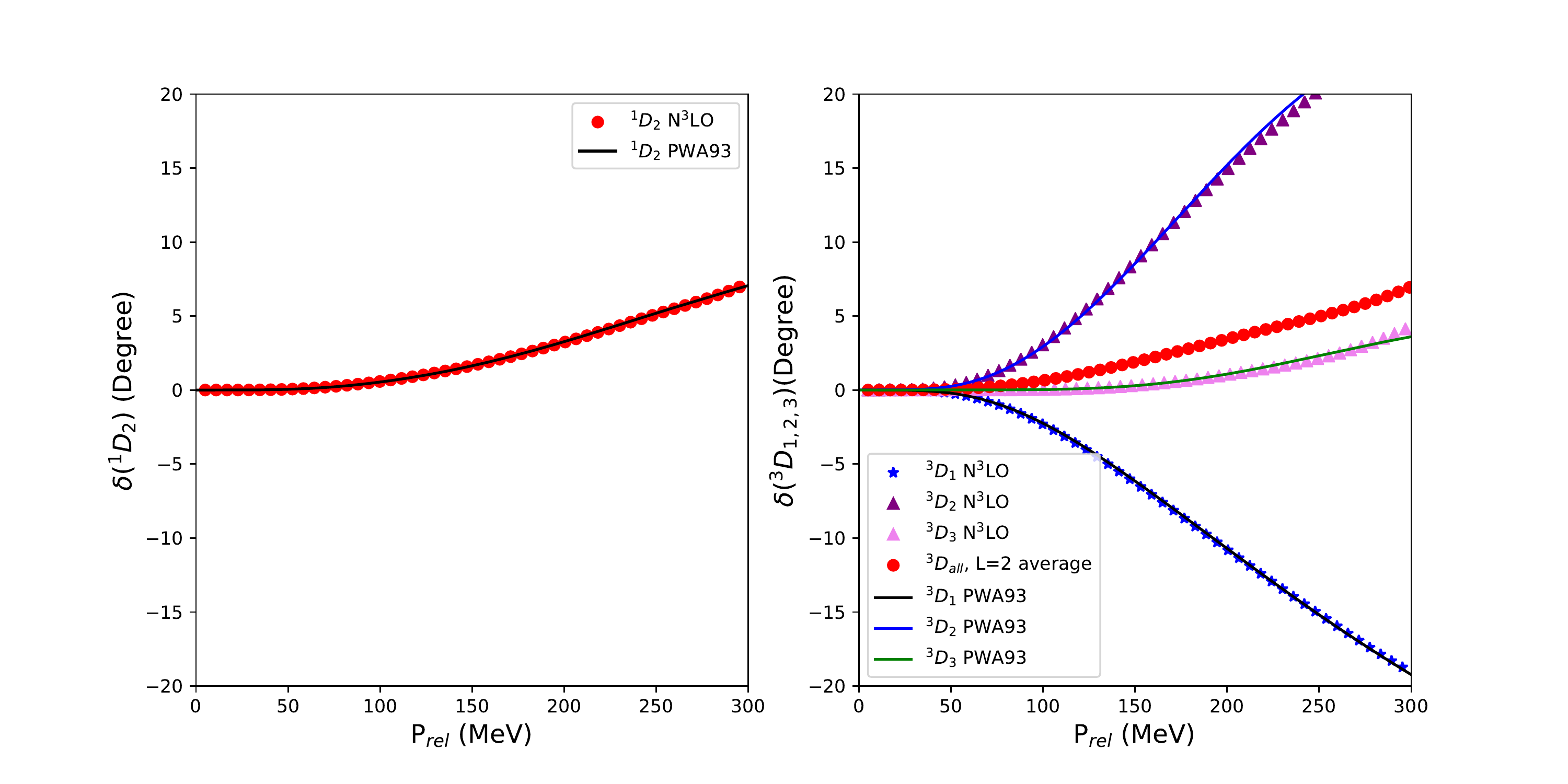}
                \label{D-wave}
        \end{figure}

Spin-isospin exchange symmetry can also be seen in the scattering phase shifts by removing the effects of the tensor interactions.  In Fig.~\ref{S-wave} we show the S-wave phase shifts obtained from lattice chiral effective field theory at next-to-next-to-next-to-leading order (N$^3$LO) for neutron-proton scattering at lattice spacing $a=1.32$~fm.  On the left panel we show the results for the $^1{\rm S}_0$ spin-singlet channel and the comparison with empirical data \cite{Stoks:1993tb}.  On the right panel we show the results for the $^3{\rm S}_1$ spin-triplet channel and the comparison with empirical data \cite{Stoks:1993tb}.  We also show the spin-triplet lattice results when the scattering is restricted to the $L=0$ orbital angular momentum channel.  This restriction removes all effects of the tensor force.  We note the good agreement between the spin-singlet partial wave and the $L=0$ spin-triplet partial wave.  This is an indication of spin-isospin exchange symmetry.
In Fig.~\ref{D-wave} we show the D-wave phase shifts at N$^3$LO for neutron-proton scattering at lattice spacing $a=1.32$~fm.  On the left panel we show the results for the $^1{\rm D}_2$ spin-singlet channel and the comparison with empirical data \cite{Stoks:1993tb}.  On the right panel we show the results for the $^3{\rm D}_1, ^3{\rm D}_2, ^3{\rm D}_3$ spin-triplet channels and the comparison with empirical data \cite{Stoks:1993tb}.  We also show the spin-triplet lattice results when restricted to the $L=2$ orbital angular momentum channel and averaged over all possible values for the total angular momentum quantum numbers $J,J_z$.  This restriction and averaging removes most though not all contributions from the tensor force as well as spin-orbit interactions.  This type of analysis with averaging over total spin $J$ was first performed in Ref.~\cite{CalleCordon:2009ps}.  We note the good agreement between the spin-singlet partial wave and the $L=2$ averaged spin-triplet partial wave.  This is again an indication of spin-isospin exchange symmetry. 

\begin{figure}[htb]
                \centering
 \caption{The $^1$S$_0$ (left) and $^3$S$_1$ (right) phase shifts for neutron-proton scattering. Dotted, dashed and dashed-dotted lines refer to the Idaho N$^3$LO \cite{Entem:2003ft}, Idaho N$^4$LO ($\Lambda = 500$~MeV) \cite{Entem:2017gor} and CD Bonn \cite{Machleidt:2000ge} potentials, respectively,  while light-shaded bands show the result of the chiral N$^4$LO$^+$ potentials from \cite{Reinert:2017usi} for the cutoff $\Lambda = 400 - 550$~MeV. For the $^3$S$_1$ channel, the results are shown when restricted to the $L=0$ orbital angular momentum channel. Solid dots correspond to the Nijmegen partial wave analysis \cite{Stoks:1993tb}.}
\includegraphics[width=12.0cm]{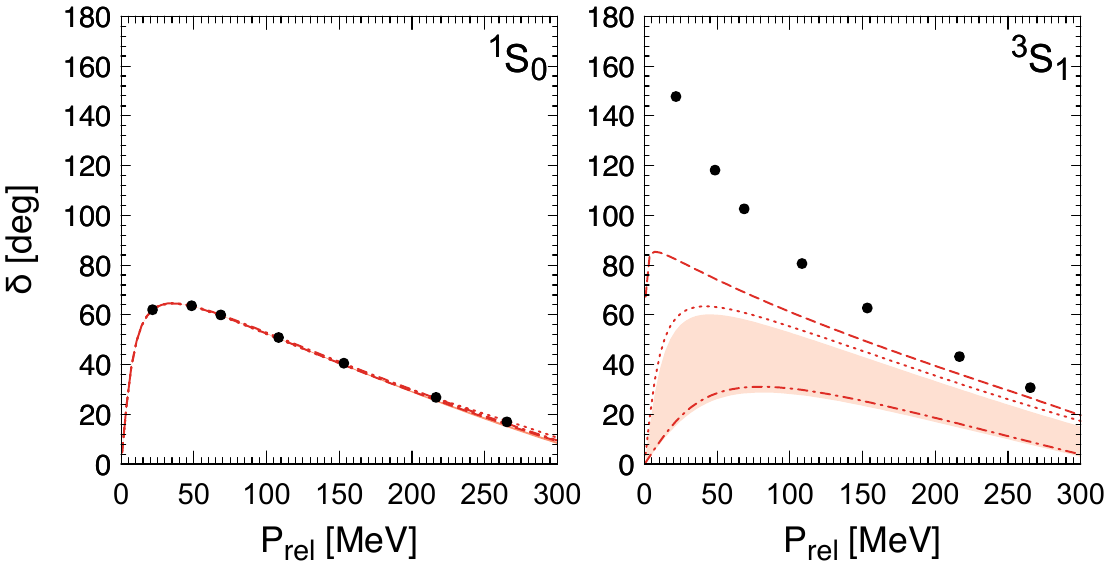}
                \label{S-wave_various}
        \end{figure}

\begin{figure}[htb]
                \centering
 \caption{The $^1$D$_2$ (left) and spin-triplet D-wave (right) phase shifts for neutron-proton scattering. Dotted, dashed and dashed-dotted lines refer to the Idaho N$^3$LO \cite{Entem:2003ft}, Idaho N$^4$LO ($\Lambda = 500$~MeV) \cite{Entem:2017gor} and CD Bonn \cite{Machleidt:2000ge} potentials, respectively,  while light-shaded bands show the result of the chiral N$^4$LO$^+$ potentials from \cite{Reinert:2017usi} for the cutoff $\Lambda = 400 - 550$~MeV. Blue lines and blue light-shaded bands show the results for the $^3$D$_1$, $^3$D$_2$ and $^3$D$_3$ phase shifts. For the spin-triplet case, red lines and red light-shaded bands show results when restricted to the $L=2$ orbital angular momentum channel and averaged over all possible values for the total angular momentum quantum numbers $J$, $J_z$. Solid dots correspond to the Nijmegen partial wave analysis \cite{Stoks:1993tb}.}
\includegraphics[width=12.0cm]{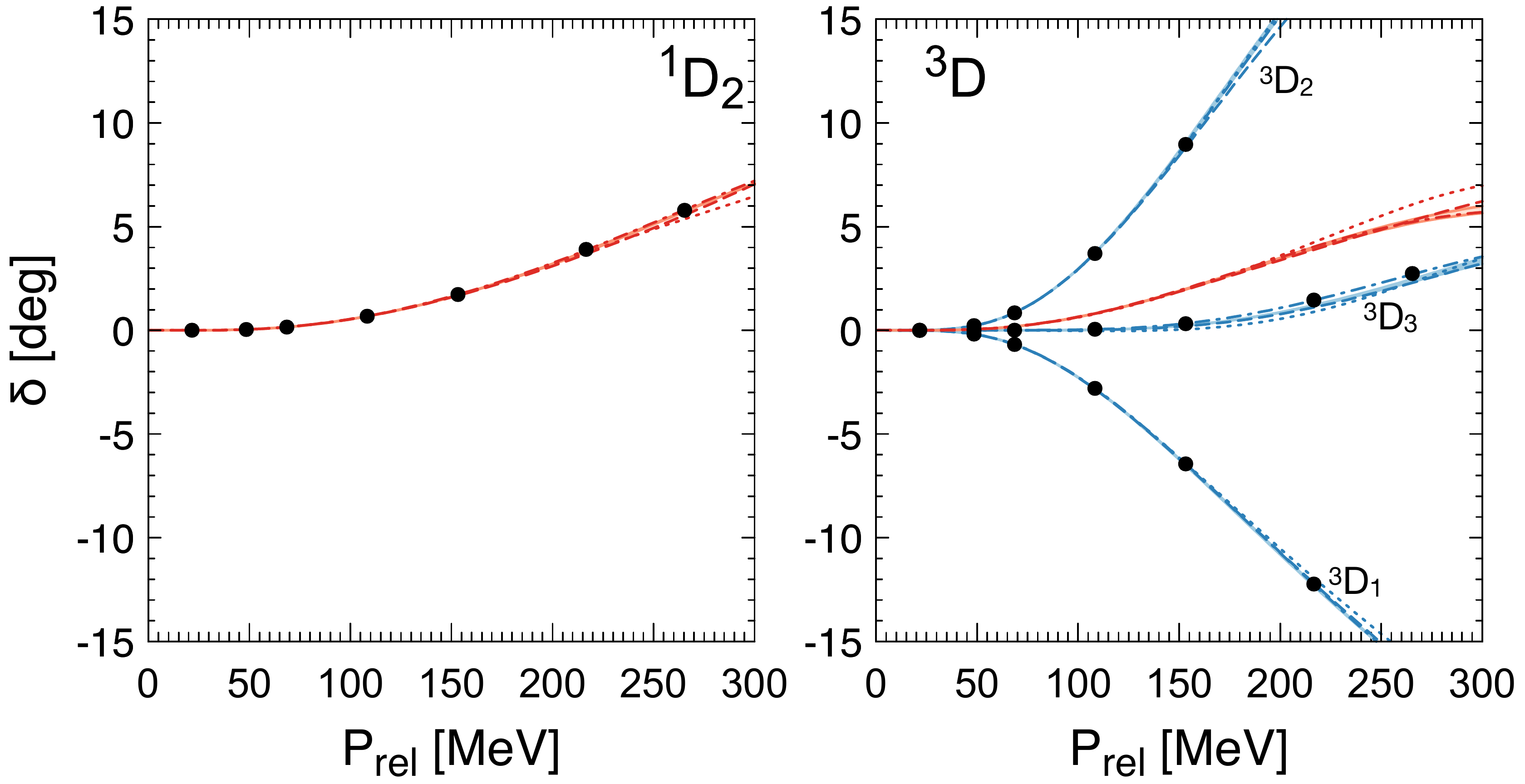}
                \label{D-wave_various}
        \end{figure}

In Figs.~\ref{S-wave_various}, \ref{D-wave_various} we show analogous results for the CD Bonn potential \cite{Machleidt:2000ge} and 
a range of two-nucleon potentials 
derived in chiral effective field theory \cite{Reinert:2017usi,Entem:2003ft,Entem:2017gor}. For the S-waves, the spin-isospin symmetry appears to be fulfilled best for the Idaho N$^3$LO and the softest version of the chiral N$^4$LO$^+$ potentials from \cite{Reinert:2017usi} corresponding to $\Lambda = 400$~MeV. In agreement with the arguments provided in our paper, the harder interactions like the CD Bonn potential and the $\Lambda=550$~MeV version of the chiral N$^4$LO$^+$ potential from Ref.~\cite{Reinert:2017usi} show a larger amount of symmetry violation.  For the D-waves, the situation is very similar to that observed for the lattice interactions. 

\subsection{Lattice interactions}
We briefly describe the lattice interactions used in this work, as there are some differences from Ref.~\cite{Li:2018ymw}.  We take the spatial lattice spacing to be $a=1/(150~{\rm MeV})=1.32$~fm.
For the free Hamiltonian we use an $O(a^{16})$-improved lattice action as defined in Ref.~\cite{Lee:2008fa,Lahde:2019npb}, meaning that the lattice discretization errors start at $O(a^{18})$.
For the leading-order interaction we use an SU(4)-invariant short-range operator $V_0$ as described in Eqs.~(24,~25) of Ref.~\cite{Li:2018ymw} as well as the one-pion exchange interaction as defined in Eqs.~(50--53) of Ref.~\cite{Li:2018ymw}.  But instead of Eq.~(52) of Ref.~\cite{Li:2018ymw}, we use 
$Q(q_S)=q_S$, where $q_S$ is the momentum along spatial direction $S=1,2,3$.  We use the nonlocal smearing parameter $s_{NL} = 0.43$, local smearing parameter $s_L = 0.074$, and coefficient $C_0 = -0.0115$ in lattice units. 
In this work we do not include the two-pion exchange interaction.

For the short-range interactions, we follow the same lattice construction based on solid spherical harmonics as described in Eqs.~(13--14) and implemented in Eqs.~(22--23) and Eqs.~(26--47) of Ref.~\cite{Li:2018ymw}.  But instead of the smeared creation and annihilation operators $\hat{a}_{i,j}^{s_{NL}\dag}({\bf n})$ and $\hat{a}_{i,j}^{s_{NL}}({\bf n})$ defined using nearest-neighbor lattice sites in Ref.~\cite{Li:2018ymw}, we use Gaussian functions in momentum space, $F({\bf q}) = \exp\left(-\frac{{\bf q}^2}{2\Lambda_{G}^2}\right)$ with $\Lambda_{G} = 400$~MeV.  Instead of $\hat{a}_{i,j}^{s_{NL}\dag}({\bf n})$, we use the Gaussian-smeared creation operator
\begin{eqnarray}
\frac{1}{L^3}\sum_{{\bf n}^\prime} \sum_{\bf q} F({\bf q}) \exp\left[i {\bf q} \cdot ({\bf n}^\prime - {\bf n}) \right]\hat{a}_{i,j}^{\dag}({\bf n}^\prime),
\end{eqnarray}
and instead of $\hat{a}_{i,j}^{s_{NL}}({\bf n})$, we use the Gaussian-smeared annihilation operator
\begin{eqnarray}
\frac{1}{L^3}\sum_{{\bf n}^\prime} \sum_{\bf q} F({\bf q}) \exp\left[-i {\bf q} \cdot ( {\bf n}^\prime - {\bf n} )\right]\hat{a}_{i,j}({\bf n}^\prime).
\end{eqnarray}

\section{Momentum and energy scales in the  large-$N_c$ limit}
The QCD scale parameter $\Lambda_{\rm QCD}$ is chosen to be independent of $N_c$. Similarly, the RMS radius of the nucleon and the chiral symmetry breaking scale, as estimated by the rho meson mass, are also independent of $N_c$. The power counting of the large-$N_c$ potential assumes that the nucleon momenta do not scale with $N_c$.  Since the mass of the nucleon scales as $N_c \Lambda_{\rm QCD}$, this implies that the nucleon velocities are $O(1/N_c)$.

Another energy scale of relevance is the mass difference between the Delta baryon and nucleon, $m_\Delta-m_N$.  This mass splitting is proportional to $\Lambda_{\rm QCD}/N_c$ and vanishes in the large-$N_c$ limit \cite{Dashen:1994qi}.  Since an effective theory containing only nucleons will not properly describe the physics above the Delta production threshold, we deduce that the momentum cutoff should be less than $\sqrt{2m_N(m_\Delta-m_N)}$.  This bound on the cutoff is independent of $N_c$, and for physical values of the Delta and nucleon masses corresponds to a maximum cutoff momentum of $740$~MeV.  We observe that our large-$N_c$ cutoff value of $\Lambda_{{\rm large-}N_c}\sim 500$~MeV is close to this upper bound.

Since the Delta production threshold is restricting the relevant energy window, it is clear that the large-$N_c$ effective theory for nucleons applies only to states near the continuum threshold.  In the large-$N_c$ limit, our effective potential will contain many deeply-bound states with binding energies that grow with $N_c$.  However, the states for which the large-$N_c$ analysis applies are those near zero energy.


\end{document}